\documentclass[fleqn,usenatbib]{mnras}

\usepackage[T1]{fontenc}

\DeclareRobustCommand{\VAN}[3]{#2}
\let\VANthebibliography\thebibliography
\def\thebibliography{\DeclareRobustCommand{\VAN}[3]{##3}\VANthebibliography}

\usepackage{graphicx}	
\usepackage{amsmath}	
\usepackage{amssymb}	
\usepackage{siunitx}
\usepackage{xspace}     
\usepackage{multirow}
\usepackage{booktabs}

\usepackage{newtxtext,newtxmath}

\newcommand{\teff}{$T_{\rm eff}$\xspace}
\newcommand{\logg}{$\log\,g$\xspace}
\newcommand{\feh}{$\rm [Fe/H]$\xspace}
\newcommand{\vmic}{$v_{\rm mic}$\xspace}
\newcommand{\vmac}{$v_{\rm mac}$\xspace}
\newcommand{\vsini}{$v \sin i_{\star}$}
\newcommand{\kms}{$\rm km\,s^{-1}$\xspace}

\title[gr8stars II : judgement day]{gr8stars II : judgement day for spectroscopic parameter model systematics}
\author[Freckelton et al.]{
Alix Violet Freckelton $^{1}$$^{\href{https://orcid.org/0009-0007-1053-0004}{\includegraphics[scale=0.5]{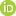}}}$\thanks{E-mail: axf859@student.bham.ac.uk},
Annelies Mortier $^{1}$$^{\href{https://orcid.org/0000-0001-7254-4363}{\includegraphics[scale=0.5]{orcid.jpg}}}$,
Megan Bedell $^{2}$ $^{\href{https://orcid.org/0000-0001-9907-7742}{\includegraphics[scale=0.5]{orcid.jpg}}}$,
Michael Cretignier $^{3}$ $^{\href{https://orcid.org/0000-0002-2207-0750}{\includegraphics[scale=0.5]{orcid.jpg}}}$,
\newauthor 
Jared R. Kolecki $^{4}$ $^{\href{https://orcid.org/0000-0002-6267-0849}{\includegraphics[scale=0.5]{orcid.jpg}}}$, 
Andreas J. Korn $^{5}$ $^{\href{https://orcid.org/0000-0002-3881-6756}{\includegraphics[scale=0.5]{orcid.jpg}}}$,
S\'ergio G. Sousa $^{6,16}$ $^{\href{https://orcid.org/0000-0001-9047-2965}{\includegraphics[scale=0.5]{orcid.jpg}}}$,
Maria Tsantaki $^{7}$ $^{\href{https://orcid.org/0000-0002-0552-2313}{\includegraphics[scale=0.5]{orcid.jpg}}}$,
John M. Brewer $^{8}$ $^{\href{https://orcid.org/0000-0002-9873-1471}{\includegraphics[scale=0.5]{orcid.jpg}}}$,
\newauthor
Lars A. Buchhave $^{9}$ $^{\href{https://orcid.org/0000-0003-1605-5666}{\includegraphics[scale=0.5]{orcid.jpg}}}$,
Guy R. Davies $^{1}$ $^{\href{https://orcid.org/0000-0002-4290-7351}{\includegraphics[scale=0.5]{orcid.jpg}}}$,
J. I. Gonz\'alez Hern\'andez $^{10,11}$ $^{\href{https://orcid.org/0000-0002-0264-7356}{\includegraphics[scale=0.5]{orcid.jpg}}}$,
Sam Morrell $^{12}$ $^{\href{https://orcid.org/0000-0001-6352-5312}{\includegraphics[scale=0.5]{orcid.jpg}}}$,
\newauthor
Martin B. Nielsen $^{1}$ $^{\href{https://orcid.org/0000-0001-9169-2599}{\includegraphics[scale=0.5]{orcid.jpg}}}$,
Vera Maria Passegger $^{13,10,11,14}$ $^{\href{https://orcid.org/0000-0002-8569-7243}{\includegraphics[scale=0.5]{orcid.jpg}}}$,
Andreas Quirrenbach $^{15}$, 
Arpita Roy $^{16}$ $^{\href{https://orcid.org/0000-0001-8127-5775}{\includegraphics[scale=0.5]{orcid.jpg}}}$,
\newauthor
Nuno C. Santos $^{6,17}$,
A. Su\'arez Mascare\~no $^{10,11}$ $^{\href{https://orcid.org/0000-0002-3814-5323}{\includegraphics[scale=0.5]{orcid.jpg}}}$,
Christopher Allan Watson $^{18}$  $^{\href{https://orcid.org/0000-0002-9718-3266}{\includegraphics[scale=0.5]{orcid.jpg}}}$,
Lily L.\ Zhao $^{19}$ $^{\href{https://orcid.org/0000-0002-3852-3590}{\includegraphics[scale=0.5]{orcid.jpg}}}$ \thanks{NASA Sagan Fellow}
\\
$^{1}$ School of Physics \& Astronomy, University of Birmingham, Edgbaston, Birmingham, B15 2TT, UK \\
$^{2}$ Center for Computational Astrophysics, Flatiron Institute, 162 5th Avenue, New York, NY 10010, USA \\
$^{3}$ Department of Physics, University of Oxford, OX13RH Oxford, UK \\
$^{4}$ Department of Physics and Astronomy, University of Notre Dame, Notre Dame, IN 46556, USA \\
$^{5}$ Division of Astronomy and Space Physics, Department of Physics and Astronomy, Uppsala University, Box 516, SE-75120 Uppsala, Sweden \\
$^{6}$  Instituto de Astrof\'{\i}sica e Ci\^encias do Espa\c co, Universidade do Porto, CAUP, Rua das Estrelas, 4150-762, Porto, Portugal \\
$^{7}$ INAF -- Osservatorio Astrofisico di Arcetri, Largo E. Fermi 5, 50125 Firenze, Italy \\
$^{8}$ Department of Physics and Astronomy, San Francisco State University, 1600 Holloway Avenue, San Francisco, CA 94132, USA \\
$^{9}$ DTU Space, Technical University of Denmark, Elektrovej 328, DK-2800 Kgs. Lyngby, Denmark \\
$^{10}$ Instituto de Astrof{\'\i}sica de Canarias, E-38205 La Laguna, Tenerife, Spain \\
$^{11}$ Universidad de La Laguna, Dept. Astrof{\'\i}sica, E-38206 La Laguna, Tenerife, Spain  \\
$^{12}$ Department of Physics and Astronomy, University of Exeter, Exeter, EX4 4QL, UK \\
$^{13}$ Subaru Telescope, National Astronomical Observatory of Japan, 650 North A‘ohoku Place, Hilo, HI 96720, USA \\
$^{14}$ Hamburger Sternwarte, Gojenbergsweg 112, D-21029 Hamburg, Germany \\
$^{15}$ Landessternwarte, Zentrum f\"ur Astronomie der Universit\"at Heidelberg, 69117 Heidelberg, Germany \\
$^{16}$ Astrophysics \& Space Institute, Schmidt Sciences, New York, NY 10011, USA \\
$^{17}$ Departamento de Fisica e Astronomia, Faculdade de Ciencias, Universidade do Porto, Rua do Campo Alegre, 4169-007 Porto, Portugal\\
$^{18}$ Astrophysics Research Centre, School of Mathematics and Physics, Queen’s University Belfast, Belfast, BT7 1NN, UK \\
$^{19}$ Department of Astronomy \& Astrophysics, University of Chicago, Chicago, IL, USA \\
\\
}

\date{Accepted XXX. Received YYY; in original form ZZZ}

\pubyear{2025}

\begin{document}
\label{firstpage}
\pagerange{\pageref{firstpage}--\pageref{lastpage}}
\maketitle

\begin{abstract}
Many areas of astrophysics, including exoplanetary studies, rely on precise and accurate stellar parameters. This demands that uncertainties on these parameters truly reflect all biases and systematics. Within this second work of the \texttt{gr8stars} collaboration, we take a set of 585 bright FGK dwarfs with high resolution, high signal-to-noise ratio spectra from the SOPHIE spectrograph. We determine stellar effective temperature, surface gravity, and metallicity using five different spectroscopic methods for each star, with an additional method used for comparisons. We find a typical scatter of 76 K in \teff, 0.14 dex in \logg, and 0.07 dex in \feh. These deviations are significantly larger than the average precision error on these parameters. We furthermore use isochrone fitting to determine mass, radius, and age for all 585 stars, using input from all results. We use the radii determined by SED fitting in the first \texttt{gr8stars} paper as a comparison to our isochronal radii from this work, in addition to comparing the isochronal \logg to spectroscopic \logg. The scatter in mass and radius from the use of different spectroscopic methods is investigated and propagated to exoplanetary parameters. The induced fractional uncertainties in planetary radius ($\lesssim$ 3 \%) and mass ($\lesssim$ 5\%) are found to be below those typically found in the literature. We estimate a lower limit on planetary equilibrium temperature fractional uncertainty of $\approx$ 4\%, a noise floor that is currently not sufficiently represented in the literature.
\end{abstract}




\section{Introduction}
Solar-like stars, defined in this work as FGK main-sequence stars, play the vital role of the host stars in the search for and study of Earth-like exoplanets. To date, no `Earth twin' exoplanet -- being an Earth-like planet orbiting a solar-like star with an orbital period on the order of Earth's -- has been identified. Many studies \citep[e.g.][]{robertson2014, meunier2015, fischer2016, suarezmascareno2017, faria2020, zicher2022, almoulla2023, santos2025} demonstrate that the lack of such a discovery is likely not due to an inherent lack of Earth twins, rather instead due to limitations in the disentangling of small planetary signals from the signals of solar-like stars.

Ongoing projects, such as the EXPRES 100 Earths Survey \citep{brewer2020} on the Lowell Discovery Telescope \citep{levine2012, degroff2014} and the NEID Earth Twin Survey \citep{gupta2021, gupta2025}, in addition to upcoming missions such as the Terra Hunting Experiment (THE) \citep{hall2018} on HARPS-3 \citep{thompson2016}, \textit{PLATO} \citep{rauer2014, rauer2025}, and the Second Earth Initiative on the Second Earth Spectrograph on the MPG/ESO 2.2m telescope (2ES; \citet{sturmer2024}), aim to overcome the current hurdles in instrumentation to kickstart the discovery and characterisation of Earth twins. In the first \texttt{gr8stars} paper (\citet{freckelton2025}, henceforth GR8-1), we introduced our sample of 5645 bright, main sequence FGKM stars that form the \texttt{gr8stars} catalogue, and published homogeneously derived stellar parameters for all Northern-hemisphere targets with available high resolution spectra. 

The increasing precision of modern, high-resolution spectrographs has in turn increased the demand for precise and accurate stellar parameters. However, it has been established that quoted precision uncertainties often fail to represent the extent of systematic effects in spectroscopic analyses. There have been numerous studies that show the significant systematics introduced in stellar parameters by the use of different methodologies, model atmospheres, analysis pipelines, and line lists \citep[see e.g.][]{hinkel2014, hinkel2016, jofre2019, tayar2022}. Such systematics frequently exceed the quoted uncertainties indicating that they present a fundamental limitation in the derivation of stellar parameters. With this context in mind, we must carefully treat the definition of stellar parameter precision. Instead, agreement between independent spectroscopic methods may provide a more realistic uncertainty estimate than internal precision errors. 

The characterisation of spectroscopic systematics has been investigated through previous collaborative endeavours. Through the Gaia-ESO survey, multiple analysis pipelines were combined to derive stellar parameters with realistic uncertainties \citep[e.g.][]{smiljanic2014, worley2020, worley2024}. Studies using the \textsc{iSpec} framework investigate controlled comparison between differing techniques \citep[e.g.][]{blancocuaresma2019, casamiquela2017, casamiquela2026}. Additionally, large-scale analyses such as the \textsc{PASTEL} catalogue \citep{soubiran2016}, and investigations of homogeneous parameter determination \citep[e.g.][]{jofre2014, heiter2015} highlight how diverse literature methodologies are, and the agreement expected between them. Further to this, comparisons of spectroscopic surveys for a common set of stars \citep[e.g.][]{hegedus2023} reinforce the presence of systematic offsets between both observations and methods.

Despite the breadth of studies performed throughout the literature, gaps still remain. Particularly, studies often rely on heterogeneous data sets, differing instrumental qualities, or samples that may only partially overlap. Consequently, there is a necessity for analyses in which multiple state-of-the-art methods are applied to the same set of high-quality spectra. This allows for the isolation of differences arising solely from systematic differences in methodology. Furthermore, although spectroscopic systematics in terms of atmospheric stellar parameters has been widely discussed throughout the literature, their propagation to further derived parameters such as stellar mass, radius, and age, is not fully quantified. This is particularly important in the context of exoplanet studies, as uncertainties in stellar parameters directly propagate to the planetary mass, radius, and density.

In the \texttt{gr8stars} catalogue, we see a uniquely suited framework for such investigations. While in GR8-1 we presented homogenously derived stellar parameters for a large sample of bright FGKM dwarfs, which is essential for a variety of applications, the data set does not fully capture the systematic uncertainties introduced by modelling choices. In this second paper, the focus is on quantifying these systematics.

A subset of 585 FGK dwarfs observed solely with the SOPHIE spectrograph is employed in this work. We apply multiple independent spectroscopic analysis techniques spanning a range of methodology, model choices, and assumptions, to each of these stars. By comparing the resultant stellar atmospheric parameters, we are able to characterise the agreement between these methods and establish estimates for the scatter introduced by methodological differences. This allows for us to place reliable constraints on the limits of precision in spectroscopic analysis, and to assess the suitability of typically quoted uncertainties to reflect true levels of agreement between independent methods.

The \texttt{gr8stars} catalogue is fully described in Section \ref{gr8stars} in terms of its selection, and context in the wider fields of exoplanetary and stellar physics. Section \ref{data} describes the spectroscopic and photometric data used in this work, along with the instruments employed for the observations. Section \ref{methods} details the methods used to determine all sets of stellar parameters presented in this work, with the results and comparisons explored in Section \ref{results}. Section \ref{exoplanets} explores the effects of scatter in stellar parameters on the parameters of hosted exoplanets. We conclude our findings in Section \ref{conc}.

\section{gr8stars}
\label{gr8stars}
The \texttt{gr8stars} catalogue is a collection of 5645 bright, FGKM stars. As part of GR8-1, we published homogeneously derived atmospheric parameters for 1716 targets in the Northern sample, in addition to photometric stellar parameters for all targets in the Northern sample. We aim to assemble, and make publicly available, a full complement of uniformly formatted, high resolution spectra, in addition to homogeneously derived stellar atmospheric parameters for all 5645 stars.

As detailed in GR8-1, three requirements were used in the selection of the \texttt{gr8stars} catalogue, with the addition of a declination cut of $\geq \ang{-15}$ to select the Northern sample that GR8-1 investigates.  The first such requirement is that all of the stars are brighter than a G band magnitude of 8, meaning archival and future observations will be to a high SNR (signal-to-noise ratio) standard. The focus of \texttt{gr8stars} on FGKM dwarfs required both earlier type main-sequence stars and giants/subgiants to be excluded. We achieved these exclusions by requiring the Gaia BP-RP colour to be at least 0.6. We used the Dartmouth isochrones and mass tracks \citep{dotter2008} to ensure our catalogue does not include evolved stars, as detailed in GR8-1.

GR8-1 published the first data set as part of \texttt{gr8stars}, the main product of which was the uniformly-formatted spectra and homogeneous stellar atmospheric parameters\footnote{\url{https://zenodo.org/records/15441644}}. Effective temperature (\teff), surface gravity (\logg), metallicity (\feh), microturbulent velocity (\vmic), macroturbulent velocity (\vmac), and projected rotational velocity (\vsini) were provided as part of the GR8-1 data set for all stars with available spectra. The \texttt{gr8stars} catalogue provides a robust and extensive resource for the stellar and exoplanet communities, with plans to extend to all targets in the catalogue through period spectroscopic archive searches and observing proposals.

\section{data}
\label{data}
As described in GR8-1, spectra were taken with 6 different high-resolution spectrographs as part of \texttt{gr8stars}. For our method comparison in this work, we use only the spectra from the SOPHIE spectrograph. This was done to ensure no scatter is introduced by the use of different instruments. The SOPHIE spectrograph covers the most stars in the current version of the catalogue. We additionally removed spectroscopic binaries, variable stars (as per their main SIMBAD object type), and targets with \vsini, as measured by \texttt{PAWS} \citep{freckelton2024} in GR8-1, larger than 10~\kms. These cuts were made to ensure we limit the amount of blended spectral lines, which severely impact the equivalent widths (EW) method, and spectral features induced by variability.

The SOPHIE spectrograph (Spectrographe pour l’Observation des Ph\'enom\'enes des Intérieurs stellaires et des Exoplan\`etes), on the 1.93 m reflector telescope at the Observatoire Haute Provence, has a wavelength coverage of 387.2 nm to 694.3 nm \citep{perruchot2008, bouchy2009}. As part of \texttt{gr8stars}, we use only observations conducted in the High Resolution (HR) mode, with a resolving power of $R \approx$ 75\,000. Of the 1286 targets observed with SOPHIE in the \texttt{gr8stars} catalogue, 585 can be classified as single, slow-rotating stars by using the SIMBAD object classifications and limiting \vsini to 10~\kms as measured by \citet{freckelton2025}. This is the subset analysed in this paper.
 
\section{Stellar Analysis Methods}
\label{methods}
\subsection{Spectroscopic Methods}
Within this work, we employ six different state-of-the-art spectroscopic analysis methods. Reflecting the variety in analysis methods in the literature, these six span not only different methodologies, but also differing atmospheric models, spectral line lists, and assumptions.
A summary of the methods and their properties can be found in Table \ref{spectro_methods}.

\begin{table*}
    \centering
    \begin{tabular}{|p{0.1\linewidth}|p{0.15\linewidth}|p{0.15\linewidth}|p{0.15\linewidth}|p{0.15\linewidth}|p{0.15\linewidth}|}
    \hline
    Analysis Code & Method Type & Free Parameters & Radiative Transfer Code & Line List & Model Atmosphere \\
    \hline
    \textsc{ARES+MOOG} & Equivalent Widths & \teff, \logg, \feh, \vmic  & \textsc{MOOG} & \citet{Sousa2008}, \citet{Tsantaki2013} & Kurucz \citep{Kurucz1993} \\
    \textsc{FASMA} & Spectral Synthesis & \teff, \logg, \feh, \vsini  & \textsc{MOOG} & \citet{Tsantaki2018} & MARCS \\
    \textsc{Metal Pipe} & Spectral Synthesis & \teff, \logg, \feh, \vsini & \textsc{MOOG}  & \textsc{Linemake} \citep{placco2021} & Phoenix \\
    \textsc{PAWS} & Equivalent Width + Spectral Synthesis & \teff, \logg, \feh, \vmic, \logg & \textsc{WIDTH}, \textsc{SPECTRUM} & \textsc{SPECTRUM} & ATLAS9 \\
    \textsc{webSME} & Spectral Synthesis & \teff, \logg, \feh, \vmic, \vmac, \vsini & \citet{wehrhahn2023} & Gaia-ESO & MARCS \\
    \textsc{YARARA} & Equivalent Widths and Machine Learning & \teff, \logg, \feh & N/A & \citet{Cretignier(2024)} & N/A \\
    \hline
    \end{tabular}
    \caption{A summary of the key components of each spectral analysis code employed for this work, in the configurations used here. }
    \label{spectro_methods}
\end{table*}

\subsubsection{ARES+MOOG}
The atmospheric stellar parameters (\teff, \logg, \feh and, \vmic) were derived using the ARES+MOOG methodology described in \citet{Santos2013,Sousa2014,Sousa2021}. For this we used ARES v2\footnote{ARES v2 can be downloaded at \url{https://github.com/sousasag/ARES}} \citep{Sousa2007, Sousa2015} to consistently measure the EW of selected iron lines for each stellar spectrum, across the entire wavelength range. For this, we used the iron line list presented in \citet{Sousa2008}, however, when we find a temperature that is lower than 5200\,K, we switch the list of lines to one that is more appropriate for cooler stars \citep[][]{Tsantaki2013}. The best spectroscopic parameters are found by converging into ionisation and excitation equilibrium. In this process, it is used for a grid of Kurucz model atmospheres \citep{Kurucz1993} and the radiative transfer code MOOG \citep{Sneden1973}. We also derived a more accurate trigonometric surface gravity using the {\it Gaia} DR3 data following the same procedure as described in \citet{Sousa2021}.

\subsubsection{FASMA}

FASMA\footnote{\url{https://github.com/MariaTsantaki/FASMA-synthesis}} \citep{Tsantaki2018, Tsantaki2020} uses the radiative transfer code MOOG (version 2019, \citealt{Sneden1973}) to compute synthetic spectra on-the-fly and determine stellar parameters through a Levenberg–Marquardt, non-linear, least-squares optimisation. The adopted line list is mostly comprised of iron lines from \cite{Tsantaki2018}, in the wavelength regions 5399 -- 5619 \AA\xspace and 6470 -- 6790 \AA. The atomic line data, namely the oscillator strengths, were obtained from the line list of the {\em Gaia}-ESO survey \citep{heiter2021}. For the rest of the lines, we kept the atomic data obtained either from the VALD database \citep{Ryabchikova2015} or calibrated empirically from observed spectra (of the Sun and Arcturus, see \citealt{Tsantaki2018} for details). The damping parameters are based on the ABO theory \citep{Barklem2000} when available, or in any other case, we used the Unsold approximation \citep{unsold1955}. FASMA performs local normalisation for the adopted intervals and a cleaning process for cosmic rays before matching with the synthetic spectrum. We used the standard solar abundances from \cite{Asplund2009} internally adopted by MOOG, except for iron, for which we adopted A(Fe) = 7.45\,dex. The model atmospheres are the grids of the MARCS libraries \citep{Gustafsson2008}. The uncertainties on the stellar parameters are derived from the covariance matrix constructed by the non-linear least-squares fit. FASMA delivers \teff, \logg, \feh, and \vsini while microturbulence and macroturbulence are refined in a second iteration based on empirical calibrations from \cite{Tsantaki2013} and \cite{Doyle2014}, respectively.

\subsubsection{Metal Pipe}

Metal Pipe \citep[ Beta v1.5.1\footnote{\url{https://github.com/kolecki4/Metal-Pipe}},][]{Kolecki+2026} is a C++ and Python framework that derives stellar abundances using line-by-line spectral synthesis. To get stellar parameters (\teff and \logg), it fits a star's photometry to MIST isochrone models \citep{dotter2016} of a given input metallicity (to start, Metal Pipe assumes a metallicity of [M/H] = 0). It then interpolates an appropriate model atmosphere from a grid of Phoenix models \citep{Husser+2013}. Using this model atmosphere, MOOG \citep[version \texttt{nov2019,}][]{Sneden1973} is called to generate synthetic spectral lines, which are fit to the observed stellar spectra using a weighted $\chi^2$ minimisation algorithm.  Metal Pipe operates on the entire wavelength range of the spectra.

First, Metal Pipe fits only iron lines. If the output metallicity (derived from synthetic spectral fitting of iron lines) does not match the input metallicity (the metallicity of the model atmosphere), the process iterates. Metal Pipe interpolates a new model atmosphere with a metallicity matching the last output value (i.e. the input metallicity of the next iteration is based on the output metallicity from the previous iteration). Again, Metal Pipe fits the iron lines to get an output metallicity and tests whether it is consistent with the input value. This iteration continues until convergence is reached.

Once [Fe/H] is converged, it fits alpha-enhancement ([$\alpha$/Fe]), using a similar method with calcium and titanium lines. Once both [Fe/H] and [$\alpha$/Fe] are converged, it fits a number of other elements as specified by the user. At the time of writing, Metal Pipe supports fitting for abundances of C, O, Na, Mg, Al, Si, S, Ca, Ti, and Fe. 

\subsubsection{PAWS}

Atmospheric parameters were determined as part of GR8-1 using the \texttt{PAWS} pipeline, introduced by \citet{freckelton2024}.  The stellar parameters presented in this work as derived by \textsc{PAWS} are directly taken from GR8-1. \texttt{PAWS} uses the \texttt{iSpec} framework \citep{blancocuaresma2014, blancocuaresma2019} to employ both the curve-of-growth EW and spectral synthesis methods. For both methods, we used the ATLAS9 model atmospheres \citep{kurucz2005}. Initial stellar parameters (\teff, \logg, \feh, \vmic) were determined using the EW method with the \texttt{WIDTH} radiative transfer code \citep{sbordone2004}. This set of initial parameters was used as an input to the spectral synthesis method, employing the \texttt{SPECTRUM} radiative transfer code \citep{gray1994} to determine the final stellar atmospheric parameters (\teff, \logg, \feh, \vmic, \vmac, \vsini). Spectral synthesis as part of \texttt{PAWS} covers the wavelength range 480 - 680~nm, using the \texttt{SPECTRUM} line list based on the NIST atomic database \citep{ralchenko2005}.

\subsubsection{webSME}
SME (Spectroscopy Made Easy) is a well-established radiative-transfer and fitting code originally coded in IDL and C++ \citep{valenti1996, piskunov2017}. It runs an adapted and updated version of the line-formation code SYNTH \citep{piskunov1992} on a pre-computed grid of plane-parallel or spherical MARCS model atmospheres \citep{Gustafsson2008}. 
SME's frontend and line-formation wrapper was recently recoded in Python \citep{wehrhahn2023} and is now referred to as PySME. 
\textsc{webSME}\footnote{\url{https://websme.chetec-infra.eu}} is a server-based interface to PySME that offers several modes. We employ two of \textsc{webSME}'s modes in the current work: the least-squares fitting mode and the forward-modelling mode. These, and the additional modes, are described in Puschnig et al. (A\&A, submitted). Output spectra can be downloaded as fits files for further offline comparisons/calculations.

The forward-modelling mode produces a synthetic spectrum for a given set of input parameters: spectral resolution, \teff, \logg, metallicity, \vmic, \vmac, \vsini, using one of several predefined line lists (here the atomic Gaia-ESO line list, \citet{heiter2021}) and one of several predefined solar reference compositions (here \citet{asplund2021}). As long as the stellar parameters fall within the grid of MARCS model atmospheres, any combination of parameters can be simulated. We note in passing that the grid of MARCS model atmospheres was computed assuming the solar reference abundances of \citet{grevesse2007}. By assuming \citet{asplund2021} for the line-formation abundance zero point, a small inconsistency is introduced in the calculations. 

The fitting mode allows the user to upload a normalised spectrum and choose any number of stellar parameters (see above) to be determined using a least-squares metric. For the present work, we focus on the determination of \logg\ from the Mg I$b$ lines assuming that the other stellar parameters are known (from runs with the other codes discussed in this section). When strictly fitting only this region, changes in Mg abundance can mimic \logg variations, producing a linear degeneracy -- see \citet{brewer2015}. To avoid this, we adopt a two step method using \textsc{webSME}. Initially, the full spectrum is fitted with Mg abundance free, constraining it using all spectral information. Secondly, we fix the Mg abundance and perform the synthesis again, limited to the region 5000 - 6000 \AA. This approach allows the \logg to be constrained using a region of the spectrum that contains the highly sensitive Mg I$b$ lines. 

Unless a specific Mg abundance is either determined or input, this procedure rests on the scaling of Mg with overall metallicity which works best at solar metallicity and may depart from the scaling relation by as much as 0.2\,dex for metal-poor thin-disk stars at \feh\,=$-0.5$ \citep{fuhrmann2004}. For the tests focusing on the six benchmark stars, this potential bias is considered to be of minor importance.   
The fitting mode needs to make many syntheses and thus takes significantly longer to produce output than the forward-modelling mode.

\subsubsection{YARARA}

YARARA is a post-processing methodology initially designed to extract more precise radial velocities by the analysis of spectra time-series \citep{Cretignier(2021),Cretignier(2023)}. A by-product of the code is the production of a master stellar spectrum free of most known instrumental systematics, including telluric lines. In this context, the pipeline has been recently upgraded \citep{Cretignier(2024)} in order to derive the stellar atmospheric parameters\footnote{\url{https://github.com/MichaelCretignier/SNAKY}} (namely \teff, \logg and \feh). For this work, we did not apply the YARARA corrections on SOPHIE spectra, but only used the atmospheric parameter retrieval recipe on them. 

The methodology relies on the extraction of 16 EWs from different chemical species (see Appendix E in \citet{Cretignier(2024)}) after the spectra have been continuum normalised by RASSINE \citep{Cretignier(2020b)}. These spectral lines are all encompassed by the wavelength region 4446 -- 6753 \AA. The choice of using several species was motivated by the release of some degeneracies between atmospheric parameters \citep{Gray(2005)}. A mapping function $F$ from $\mathbb{R}^{16}$ to $\mathbb{R}^3$: $$F(EW_1,EW_2,...,EW_{16}) = (\text{\teff}, \text{\logg},\text{\feh})$$ was then fitted using a non-linear approach such as machine learning regressions. In order to fit the function $F$, the full HARPS database was reduced to extract the EWs and a collection of several catalogs of stellar atmospheric parameters was obtained from different methods and instruments (see references in \citet{Cretignier(2024)}). Because the function $F$ was calibrated using HARPS spectra, we first established the scaling coefficient between the HARPS and SOPHIE EWs that accounts for the different spectral resolution of both instruments. The coefficient was obtained from a star that had been observed with both instruments. 

By definition, the YARARA predicted parameters can be understood as the average answer that would have been provided by the cited catalogues/methods. The estimated uncertainties on the different parameters were given by the test sample and are $\pm$ 70~K, $\pm$ 0.07~dex, and $\pm$ 0.07~dex on \teff, \logg and \feh respectively. As a drawback of the methodology, the machine learning regressor struggles to predict extreme values for stellar observations that were not in the training sample (such as stars cumulating anomalous values in \teff, \logg or \feh). 

\subsection{Isochrone Fitting}
\label{isochrone_method}
Using the stellar atmospheric parameters in conjunction with parallax and magnitudes, we used the \textsc{isochrones} \citep{morton2015} package to determine masses, radii and ages for the \texttt{gr8stars} targets by following the methodology outlined by \citet{mortier2020}. As inputs, we used \teff and \feh from our spectroscopic analyses, in addition to the Gaia DR3 parallax, 2MASS J, H, and K magnitudes \citep{skrutskie2006}, and AllWISE W1, W2, and W3 magnitudes \citep{cutri2013}. Note that we did not include spectroscopic \logg in this analysis. Although accurate spectroscopic \logg measurements are achievable, there are potential biases and offsets depending on the method used at a significance that would be relevant for isochrone fitting \citep[e.g.][]{torres2012, mortier2014, magrini2022}. The use of photometric magnitudes combined with parallax instead provides a more accurate constraint on stellar radius, and hence \logg. We employed the MESA Isochrones and Stellar Tracks (MIST, \citet{dotter2016}) stellar evolution model. In this work we present both the individual masses, radii, and ages from running \textsc{isochrones} on the results from each spectroscopic method, and also an inverse-variance-weighted mean of the MAP (Maximum a posteriori) values from all methods. The latter should automatically account for systematic differences between spectroscopic methods, but still relies on only one set of stellar evolution models which can have systematics of its own. As a result, this work establishes only a lower limit on the estimated systematics.

\section{Stellar Parameter Results}
\label{results}

To understand the systematic differences and accuracies of derived stellar parameters, we compare here the stellar parameters derived from the codes to each other, in addition to comparing to literature results.

\subsection{Method differences}
From all spectroscopic methods, we obtain values for \teff, \logg, and \feh, in addition to propagating these through \textsc{isochrones} to determine radius, mass, and age. Given the results of all spectroscopic methods, we can estimate the typical scatter on all six of these stellar parameters that arises due to differences in methodology and models. To do so, we calculate the median value of each parameter for each star across all sets of results. Then, for each star, we calculate the RMSD (Root Mean Square Deviation) across the different methods from this median value. The overall scatter is then taken as the median RMSD across all stars. Table \ref{tab:scatters} displays the results of this, showing the typical scatter that we find in \teff, \logg, \feh, radius, mass, and age. 

\begin{table}
    \centering
    \caption{The typical scatter found from the use of different spectroscopic methods for stellar \teff, \logg, \feh, radius, mass, and age.}
    \begin{tabular}{|c|c|}
    \hline
         Parameter & Typical Scatter  \\
         \hline
         \teff (K) & 76 \\
         \logg (dex) & 0.14 \\
         \feh (dex) & 0.07 \\
         Radius (R$_{\odot}$) & 0.02 (1.22 \%)\\
         Mass (M$_{\odot}$) & 0.05 (5.50 \%)\\
         Age (Gyr) & 2.08 ($>$ 100 \%) \\ 
         \hline
    \end{tabular}
    
    \label{tab:scatters}
\end{table}
The floors in systematic uncertainties identified in Table \ref{tab:scatters} correspond to approximately 1.22 \% in stellar radius, 5.50 \% in stellar mass, and over 100 \% in stellar age. These values are, broadly, consistent with the precision requirements for \textit{PLATO} stellar radius and mass\footnote{\url{https://platomission.com/2018/06/05/science-performance/}}. This indicates that spectroscopic method differences alone are unlikely to represent the dominant limitation in stellar mass and radius determinations for \textit{PLATO} targets. In contrast to this, the scatter in age is substantially larger, demonstrating the strong sensitivity of isochrone derived ages to scatter in the employed atmospheric parameters.

\subsubsection{Spectroscopic results}
\label{pastel}
The PASTEL catalogue \citep{soubiran2016} provides a useful resource of stellar atmospheric parameters, collating results from a variety of sources and methods. Within this work we use the 2020-01-30 version of the catalogue for comparison. We note that the \feh measurements included in the PASTEL catalogue are all directly from high resolution, high SNR spectroscopy, making them relevant comparison values. The \teff measurements are predominantly from spectroscopy and photometry, with a small number having \teff determined from interferometry. In terms of \logg, most values result from the ionisation-equilibrium and parallax methods, however the PASTEL catalogue does include \logg values determined through asteroseismology. It is unlikely that any of the stars within this work have a PASTEL \logg measured using asteroseismology, given that we focus on only FGK dwarfs.
Given that the median values of the atmospheric parameters, and their associated uncertainties, represent both the consensus and spread in values at the current state of the field, we expect our results in this work to reflect those from PASTEL. Throughout this work, the PASTEL values used for comparison are the median of all PASTEL values for the star. We interpret the inherent scatter in the PASTEL catalogue as the standard deviation present in the results collated for each target. Across all targets in this work with multiple sources of parameters, we see median standard deviations in \teff, \logg, and \feh of 52 K, 0.09 dex, and 0.04 dex respectively in the PASTEL catalogue.

\begin{figure*}
    \centering
    \includegraphics[width=\textwidth]{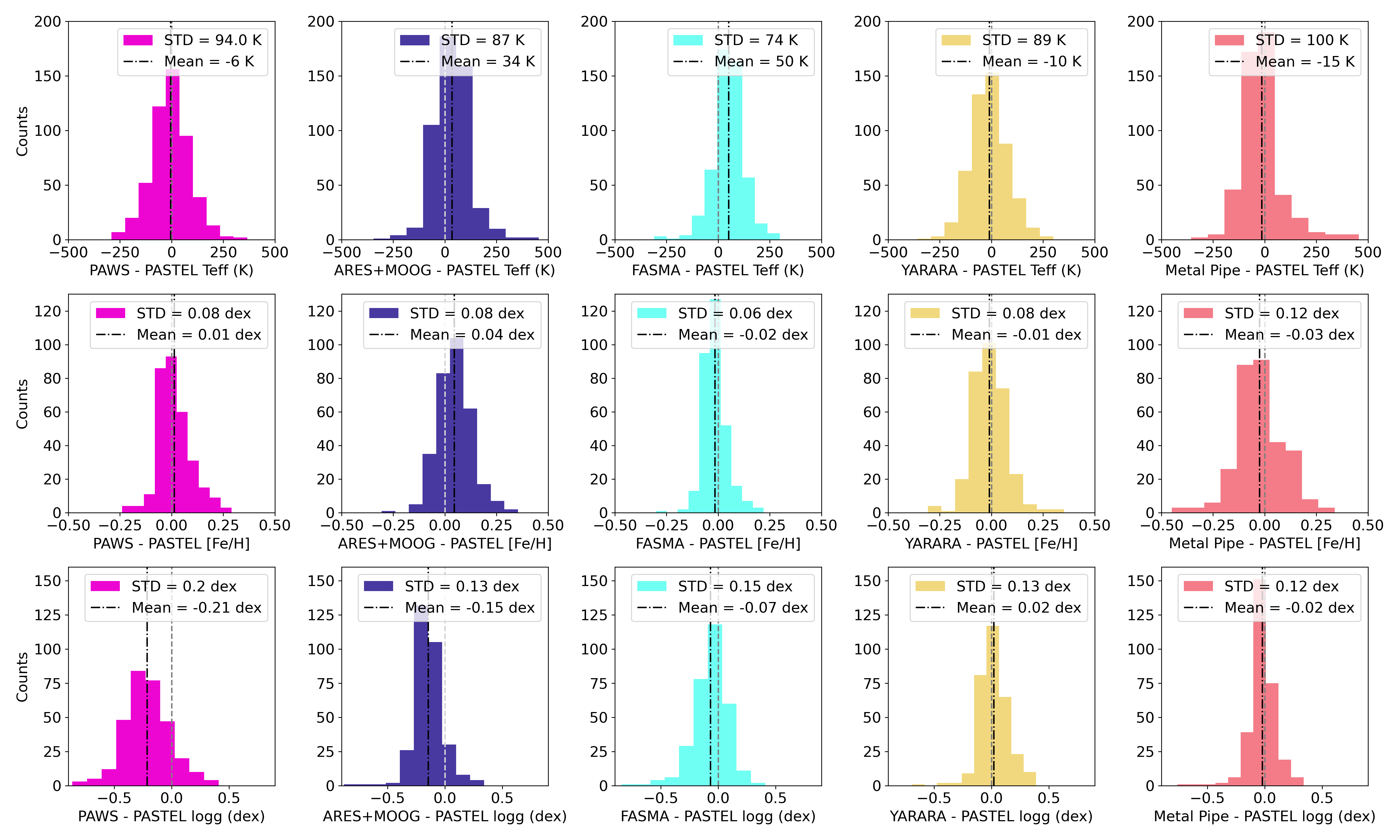}
    \caption{Histograms of the differences between results from each method and PASTEL for \teff, \feh, and \logg. The standard deviation (STD) and mean of these histograms is marked on each plot, with a grey dashed line representing 0 in each plot.}
    \label{fig:big_pastel_comp}
\end{figure*}

Figure \ref{fig:big_pastel_comp} shows histograms of the differences between atmospheric parameters in this work and those from PASTEL, split into \teff, \feh, and \logg as the top, middle, and bottom rows, respectively. For each plot, we would expect an unbiased, accurate method to produce a histogram of uncertainties that is Gaussian and centred around 0 difference, with a FWHM (Full Width at Half Maximum) reflective of the typical uncertainties in the methods. This is observed in the majority of cases in Figure \ref{fig:big_pastel_comp}, with a few notable exceptions. We show the RMSD, median difference, and median uncertainty for all results in Table \ref{tab:pastel_stats}.

Observing the difference between \teff derived in this work and that from PASTEL in Figure \ref{fig:big_pastel_comp}, the RMSD is between 89 and 102~K for all targets. This follows investigations performed by \citet{heiter2015}, in which the PASTEL catalogue was analysed for the Gaia-ESO benchmark stars. From this, it was found that for stars with three or more spectroscopic determinations of \teff showed a mean standard deviation in \teff of $\sim$ 85 K. We repeat such an investigation for the stars in this work with three or more PASTEL entries. This totals 310 stars for \teff, giving a median spread (maximum -- minimum) of 131 K. Across the 202 stars with three or more \logg and \feh entries, we see a median spread of 0.20 dex in \logg, and 0.12 dex in \feh. Comparing these to the scatter we see in our results through Figure \ref{fig:big_pastel_comp} and Table \ref{tab:pastel_stats}, we see that our scatter and that seen in PASTEL is of a similar magnitude.

\begin{figure*}
    \centering
    \includegraphics[width=\linewidth]{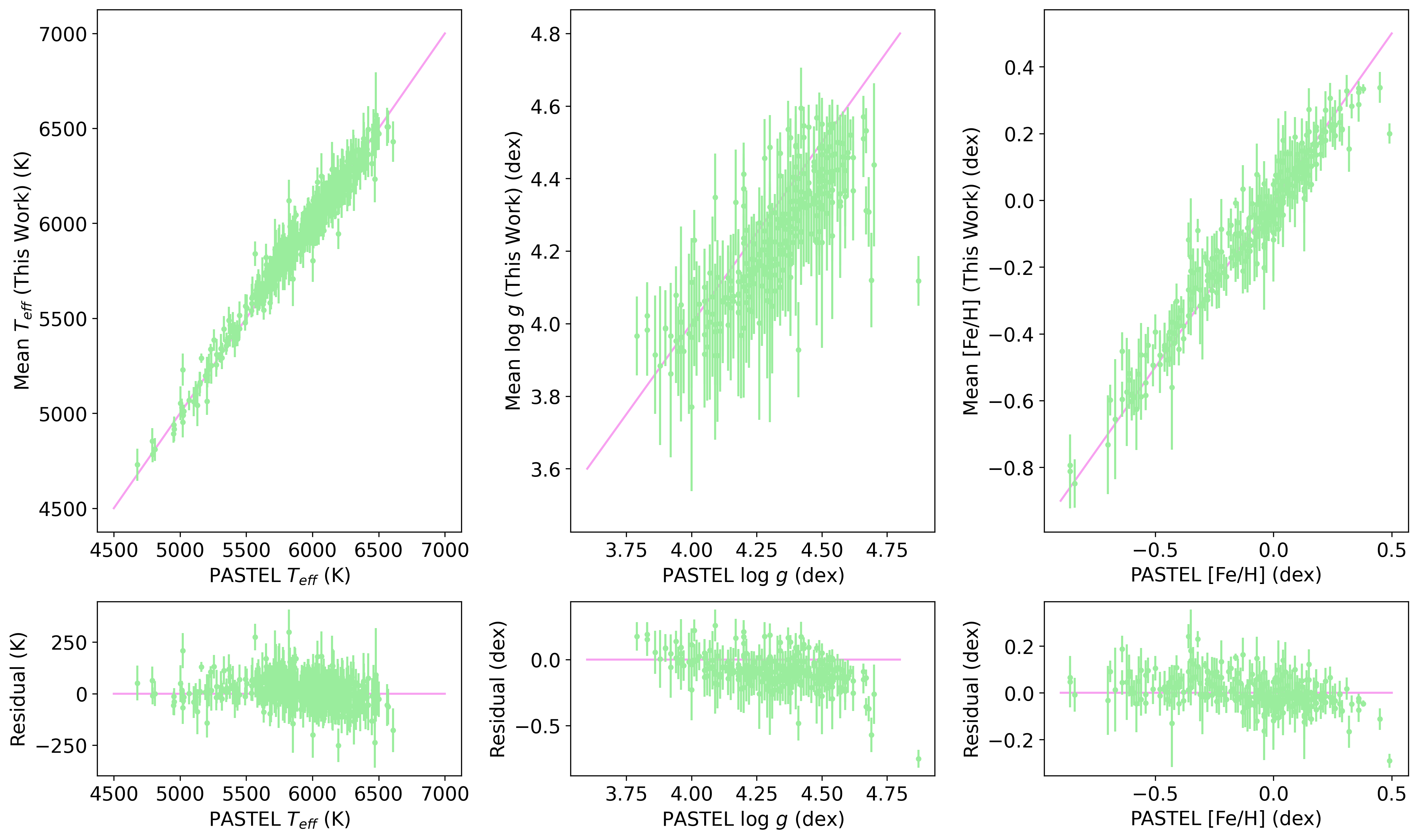}
    \caption{The mean value from all spectral analysis codes in this work plotted against the PASTEL value for \teff (left), \logg (centre), and \feh (right). The pink line in all panels represents the one-to-one relationship. Lower panels show the residuals between the data sets. Y axes errors represent the standard deviation in the results from this work.}
    \label{fig:meanmean}
\end{figure*}

In Figure \ref{fig:meanmean}, we plot the mean results from all analysis codes against the PASTEL values for \teff, \logg, and \feh, respectively. The most striking trend here is the systematically lower \logg values returned in this work compared to the PASTEL values. Given that PASTEL results come from various sources, not limited to spectroscopy, we can assume that this is a spectroscopic effect. The mean difference between our mean \logg and that from PASTEL is -0.09 dex. However, this lies within the mean standard deviation we see in our results of 0.12 dex, suggesting that uncertainties due to method-dependent scatter outweigh this systematic underestimation.

When we consider the median differences shown in Table \ref{tab:pastel_stats}, we see that \textsc{ARES+MOOG}, \textsc{FASMA}, and \textsc{YARARA} results show median differences at absolute values larger than their respective median uncertainties. This could indicate a systematic offset, however as the uncertainties are precision based, and do not account for model inaccuracies, it is likely that uncertainties adjusted for this would encompass the observed median difference. This is further supported by the fact that both \textsc{PAWS} and \textsc{Metal Pipe} have median uncertainties reflective of their scatter (RMSD), whereas scatter significantly outweighs uncertainty for \textsc{ARES+MOOG} and \textsc{FASMA}.

Regarding \feh, we can see from Table \ref{tab:pastel_stats} that the RMSD is larger than the median uncertainty for all results, indicating underestimated uncertainties in all cases. We can additionally observe that the median difference between \feh from \textsc{ARES+MOOG} and \textsc{Metal Pipe} and that from PASTEL is larger than the median uncertainty.

By far the most obvious issues in Figure \ref{fig:big_pastel_comp} lie with the \logg. Underestimation of \logg is clear from both \textsc{PAWS} and \textsc{ARES+MOOG}, with median differences of -0.22 and -0.16~dex, respectively, exceeding the median uncertainties in both cases. Both \textsc{PAWS} and \textsc{ARES+MOOG} also show RMSD at scales larger than three times the median uncertainty, strongly suggesting that uncertainties are underestimated for the resulting \logg. Both the median differences and RMSD in the results suggest \logg derived from the spectroscopic methods within this work requires larger uncertainties than purely precision based ones, to account for methodological and model differences. In terms of \teff and \feh, we see far more significant standard deviations in our work of 64 K and 0.061 dex respectively, than their mean differences from PASTEL of 11 K and 0.001 dex. It is clear that for \teff and \feh, method-to-method scatter is a significant source of uncertainty.

\begin{figure*}
    \centering
    \includegraphics[width=\textwidth]{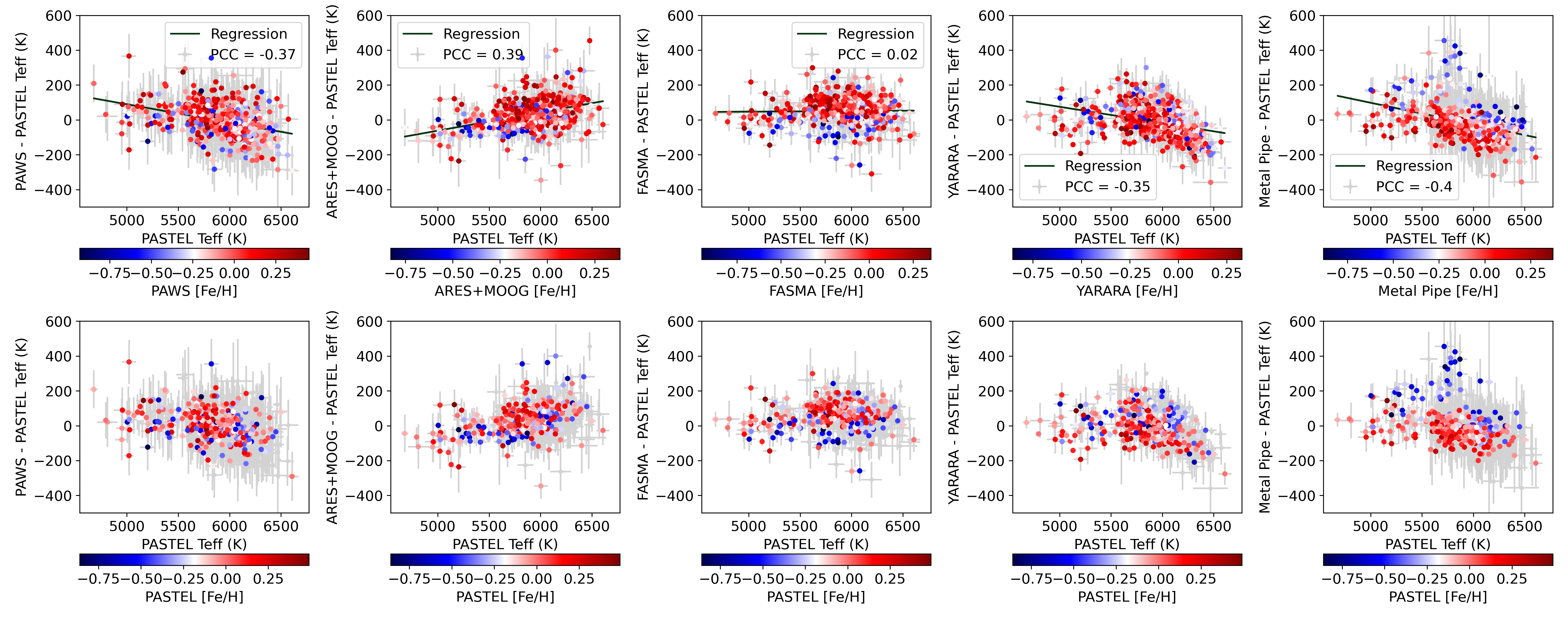}
    \caption{Differences in \teff from this work and those from PASTEL, coloured by the \feh from each method on the top row, and the PASTEL \feh on the second row. The top row additionally includes linear regression lines fitted to the data, as well as the Pearson Correlaton Coefficient (PCC) for the five sets of results.}
    \label{fig:pastel_feh}
\end{figure*}

\begin{table*}
    \centering
    \begin{tabular}{|c|c|c|c|c|c|c|c|c|c|c|c|c|c|c|c|}
    \hline
         \multirow{2}{*}{Parameter} & \multicolumn{5}{c|}{RMSD}  & \multicolumn{5}{c|}{Median Difference} & \multicolumn{5}{c|}{Median Uncertainty} \\
         \cline{2-16}
          & P & A+M  & F & Y & MP & P & A+M & F & Y & MP & P & A+M & F & Y & MP \\
          \hline \hline
          $\Delta$ \teff (K) & 94 & 94 & 90 & 89 & 101 & -8 & 34 & 51 & -14 & -27 & 105 & 30 & 26 & 70 & 93 \\
          $\Delta$ \feh & 0.08 & 0.10 & 0.06 & 0.08 & 0.12 & 0.00 & 0.04 & -0.02 & -0.02 & 0.04 & 0.04 & 0.02 & 0.03 & 0.07 & 0.01 \\
          $\Delta$ \logg & 0.29 & 0.19 & 0.17 & 0.13 & 0.12 & -0.22 & -0.16 & -0.05 & 0.01 & -0.02 & 0.12 & 0.05 & 0.09 & 0.07  & 0.04 \\
          \hline

    \end{tabular}
    \caption{The RMSD and median difference from the PASTEL values, and median uncertainties on the \teff, \feh, and \logg, for \textsc{PAWS} (P), \textsc{ARES+MOOG} (A+M), \textsc{FASMA} (F), \textsc{YARARA} (Y), and \textsc{Metal Pipe} (MP).}
    \label{tab:pastel_stats}
\end{table*}
We further investigate differences between \teff from this work and that in the PASTEL catalogue in Figure \ref{fig:pastel_feh} by plotting these against the PASTEL \teff, coloured by the \feh from each method. Each plot is repeated below coloured by the PASTEL \feh, to allow for an alternative comparison independent of method. For each comparison of \teff difference to PASTEL \teff, we calculated the Pearson correlation coefficient (PCC) to identify any trends. Out of all methods, only \textsc{FASMA} shows little-to-no correlation, with a PCC of 0.02. Additionally, \textsc{ARES+MOOG} is the only method showing a positive correlation, with a value of 0.39. whereas \textsc{PAWS}, \textsc{YARARA}, amd \textsc{Metal Pipe} have negative correlations of -0.37, -0.35, and -0.40, respectively. This evaluation suggests that \textsc{FASMA} determines \teff without bias relative to the PASTEL values across the FGK regime. Limiting the sample to solar and cooler stars, based on the PASTEL \teff, the correlations for the latter three results change significantly. For \textsc{PAWS} and \textsc{Metal Pipe}, the PCCs become -0.05 and 0.01, respectively, in the solar and cooler regime, suggesting an improved suitability here as opposed to with hotter stars. Interestingly, the \textsc{YARARA} \teff difference becomes positively correlated with PASTEL \teff at PCC of 0.22. 

Figure \ref{fig:pastel_feh} additionally shows trends with \feh. Particularly prevalent here are the results from \textsc{Metal Pipe}. Colouring the points in this plot by both the \textsc{Metal Pipe} and PASTEL \feh values reveals an influence of \feh on \teff, or vice versa. This is not unexpected, given that changes in one stellar parameter can be at least partially compensated by apparent changes in another, particularly when using small numbers of lines \citep{Gray(2005)}. High metallicity targets tend to have an overestimated \teff compared to PASTEL when determined by \textsc{Metal Pipe}, whereas lower metallicity stars see an underestimated \teff compared to PASTEL. Figure \ref{fig:pastel_feh} is replicated for \teff and \logg in Appendix \ref{srsly_name_this_properly}.
 
\subsubsection{Isochrone results}
\begin{figure*}
    \centering
    \includegraphics[width=\textwidth]{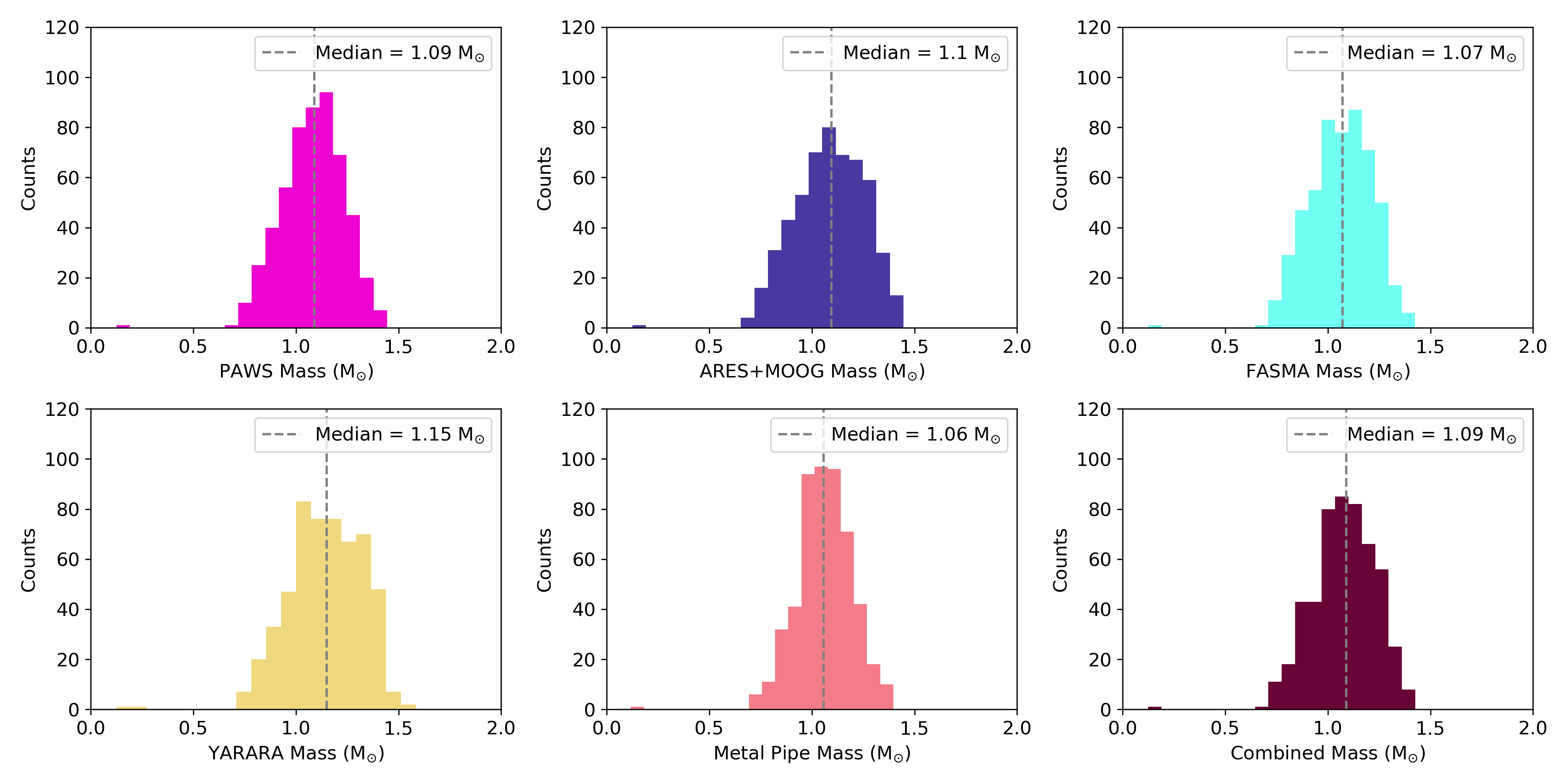}
    \caption{Histograms of the whole-sample masses obtained using \textsc{isochrones}, with \teff and \feh from \textsc{PAWS}, \textsc{ARES+MOOG}, \textsc{FASMA}, \textsc{YARARA}, \textsc{Metal Pipe}, and the combined results. Grey dashed lines indicate the median of each distribution of results.}
    \label{fig:mass_hist}
\end{figure*}

As detailed in Section \ref{isochrone_method}, we used the spectroscopic \teff and \feh, together with photometric magnitudes, to determine mass and radius for each star, amongst other parameters. Figure \ref{fig:mass_hist} shows histograms of the resulting stellar mass distributions from \textsc{PAWS}, \textsc{ARES+MOOG}, \textsc{FASMA}, \textsc{YARARA}, and \textsc{Metal Pipe}, in addition to the mass distribution from the combined isochronal posterior. This shows that our sample is weighted towards stars heavier than solar, with all distributions peaking higher than solar mass. This is to be expected, given that the mass tracks used to select the whole \texttt{gr8stars} sample (see GR8-1) retain metal rich F dwarfs, however not metal poor ones. This biases the F type population, and therefore in part the entire \texttt{gr8stars} population, to heavier stars.

\begin{figure*}
    \centering
    \includegraphics[width=\textwidth]{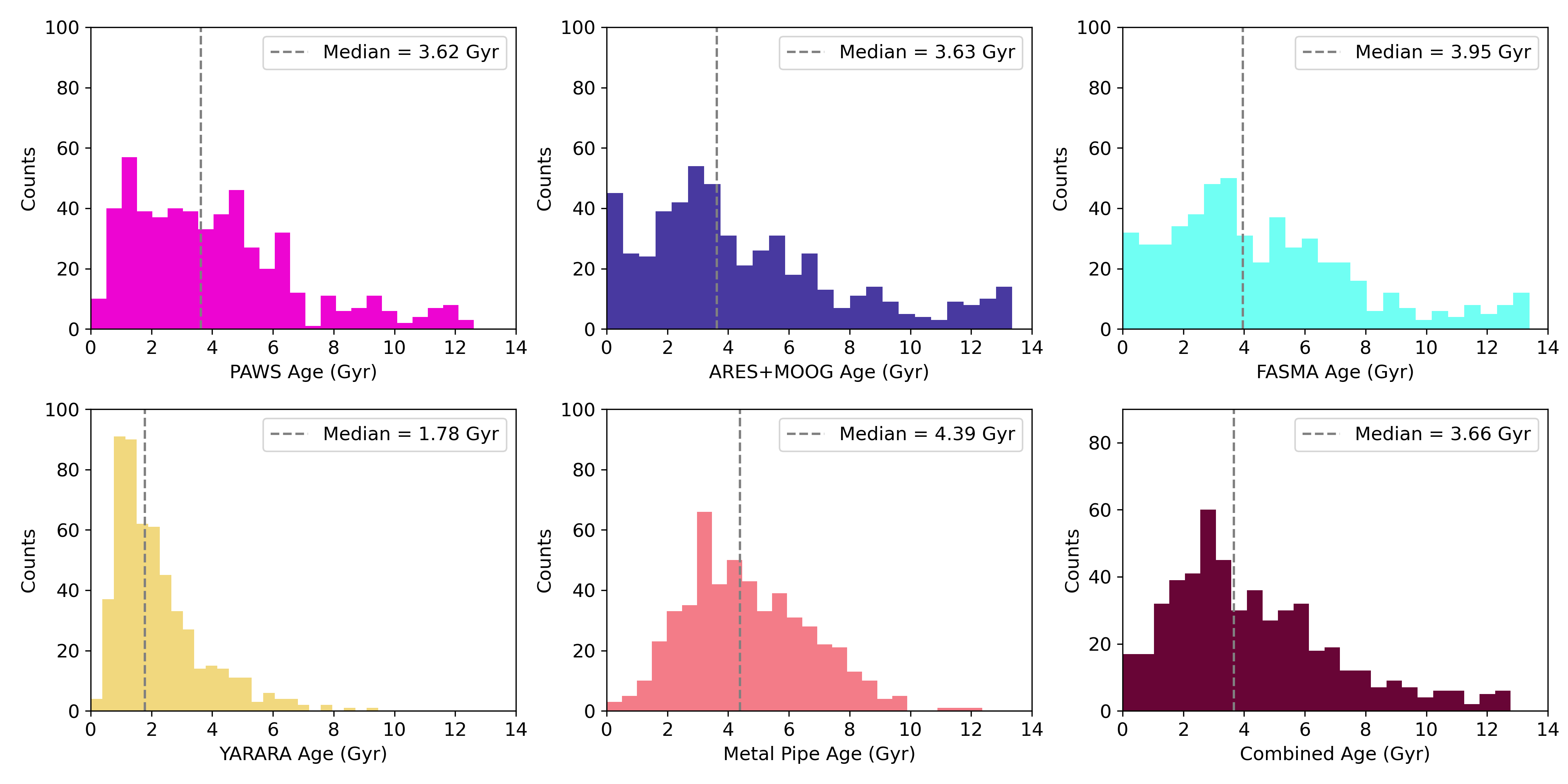}
    \caption{Distribution of stellar ages across the sample from this work. Stellar ages were determined using \textsc{isochrones}, with results from inputting spectroscopic \teff and \feh from \textsc{PAWS}, \textsc{ARES+MOOG}, \textsc{FASMA}, \textsc{YARARA}, \textsc{Metal Pipe}, and the combined results.  Grey dashed lines display the median age from each distribution of results.}
    \label{fig:age_hists}
\end{figure*}

In terms of stellar ages, typical uncertainties on the isochronal findings are on the order of $\approx$ 30\% \citep{xiang2017}, owing to the slow timescales of change of observables. Isochrones for different ages are very close for main sequence stars.
We see these uncertainties reflected in our own results from \textsc{isochrones} in this work. The distribution of stellar ages determined through \textsc{isochrones} and spectroscopic inputs is shown in Figure \ref{fig:age_hists}. The individual distributions from the use of different spectroscopic methods follow a broad shape that peaks towards a region between 1 and 5~Gyr, however the value of this peak varies significantly between the codes used. For any given star, we see a typical scatter in the determined age of 2.08~Gyr solely from the use of \teff and \feh inputs from different spectroscopic methods. 

For mass, radius, and age, all difference in results comes from the scatter in \teff and \feh from the spectroscopic methods.  As shown in Table \ref{tab:scatters}, we see scatter in mass of 0.05~M$_{\odot}$, radius of 0.02~R$_{\odot}$, and age of 2.08~Gyr. It is important to contextualise these in terms of the typical precision uncertainties we see on these parameters using \textsc{isochrones}. Across all of our results, we see a median uncertainty in stellar mass of 0.03~M$_{\odot}$, stellar radius of 0.01~R$_{\odot}$, and stellar age of 1.66~Gyr. Given that these uncertainties are lower than the scatter we observe when using spectroscopic inputs from different methods, we suggest that these do not fully reflect the true uncertainties in these parameters.

\subsection{Effective Temperatures}
\label{benchmark}
\subsubsection{Interferometry}
A useful tool for our method comparison is the availability of stellar parameters for benchmark stars. \citet{soubian2024} provide interferometric fundamental \teff and radii measurements for 192 Gaia FGK benchmark stars. While these measurements have minimal reliance on spectroscopy or evolutionary models, they do not represent complete model independence. The bolometric fluxes used are determined through SED fitting, and limb-darkening models are applied to the stellar angular diameters.
Cross-matching with \citet{soubian2024}, we find a small overlap of 6 targets that are also in our sample of 585 stars in this paper.

\begin{figure}
    \centering
    \includegraphics[width=0.5\textwidth]{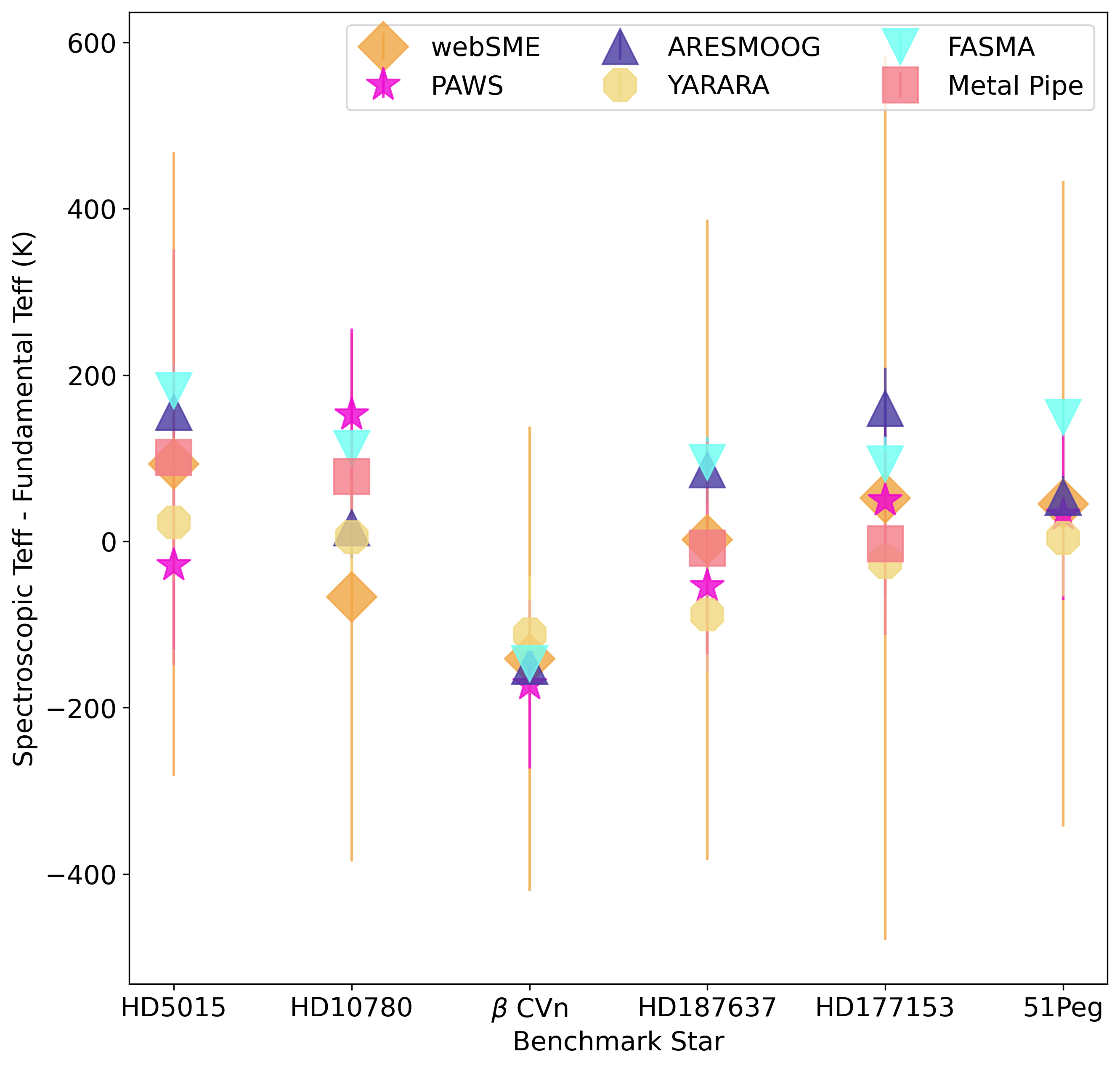}
    \caption{The difference between spectroscopic and fundamental (interferometric) \teff for each method applied in this work, for each of the 6 stars in common with \citet{soubian2024}.}
    \label{fig:benchmark}
\end{figure}

Figure \ref{fig:benchmark} displays the difference in \teff between the spectroscopic methods in this work, and the interferometry from \citet{soubian2024} for these six stars. We see overall agreement between all methods and the interferometric effective temperatures. Across all methods, the mean difference from the interferometric effective temperature is 78~K, sitting comfortably within the interferometric mean uncertainty in \teff of 135~K. Considering the differences in \teff between the methods used in this paper, we see a mean dispersion across the six stars of 211~K. \citet{soubiran2016} compare the fundamental \teff values for 151 stars to those from photometry and spectroscopy, finding a median absolute deviation of 96~K. Our results here are therefore within the broad expectations of comparing spectroscopic to interferometric \teff. 

An additional observation from Figure \ref{fig:benchmark} is that for $\beta$ CVn we see strong agreement between \teff results from our spectroscopic methods, however at an offset when compared to the fundamental \teff from \citet{soubian2024}. All spectroscopic results are within 60~K of each other for $\beta$ CVn, hence indicating that the spectroscopic methods agree for this star, however differ somewhat significantly compared to the interferometric \teff. In terms of other spectroscopic parameters, we see a standard deviation from the six codes of 0.17 dex in \logg, and 0.18 dex in \feh.

In performing a literature search for $\beta$ CVn, we found two other interferometric \teff measurements. \citet{boyajian2012} gives a \teff of 5653 $\pm$ 72~K, while \citet{baines2018} measures 5966 $\pm$ 117~K, both values cooler than the 6013 $\pm$ 91~K obtained from \citet{soubian2024}. Of particular note is the \teff from \citet{boyajian2012} being cooler than both other interferometric values by over 300~K. Differences in interferometric values could result from uncertainty in the derivation of angular diameter, or differing treatments of limb darkening and bolometric flux determination. Given the small sample size, it is not possible to truly disentangle systemic effects in spectroscopic methods from those that are inherent in the interferometric analysis.

\begin{table}
    \centering
    \caption{Effective temperature for $\beta$ CVn as derived from spectroscopy in this work, and from interferometry, spectroscopy, and photometry in the literature.}
    \label{tab:littyps}
    \begin{tabular}{|cc|}
    \hline
        Source & $\beta$ CVn \teff (K) \\
        \hline
        & \textit{Spectroscopy (This Work)}  \\
        \textsc{webSME} & 5872 $\pm$ 279  \\
        \textsc{ARES+MOOG} & 5863 $\pm$ 20  \\
        \textsc{FASMA} & 5866 $\pm$ 19  \\
        \textsc{PAWS} & 5841 $\pm$ 101  \\
        \textsc{YARARA} & 5901 $\pm$ 70 \\
        \textsc{Metal Pipe} & --  \\
        & \textit{Literature Interferometry}  \\
        \citet{soubian2024} & 6013 $\pm$ 91\\
        \citet{boyajian2012} & 5653 $\pm$ 72  \\
        \citet{baines2018} & 5966 $\pm$ 117  \\
        & \textit{Literature Spectroscopy}  \\
        \citet{mishenina2015} & 5897  \\
        \citet{gonzalez2010} & 5806 $\pm$ 24 \\
        \citet{dasilva2015} & 5875 $\pm$ 30 \\
        & \textit{Literature Photometry}  \\
        \citet{luck2017} & 5865  \\
        \citet{ramirez2007} & 5779  \\
        \citet{casagrande2011} & 5879 \\
        \hline

    \end{tabular}
\end{table}

Given the agreement between spectroscopic methods we observe for $\beta$ CVn, it is of interest to see if this propagates to including literature values aside from the interferometry from \citet{soubian2024}. Table \ref{tab:littyps} displays the stellar effective temperature determined as part of this work, in addition to literature values from interferometry, spectroscopy, and photometry. We see a general agreement between our results and those of literature spectroscopy and photometry that suggest the star is cooler than the interferometric methods of \citet{soubian2024} and \citet{baines2018}, however hotter than that from \citet{boyajian2012}. 

\subsubsection{SED}
As part of GR8-1, photometric \teff and radii were obtained following the spectral energy distribution (SED) fitting methodology first applied by \citet{morrell_naylor2019, morrell_naylor2020, morrell2025} -- see GR8-1 for a detailed description of how this was used. Comparing the results from this work to these photometrically derived values allows for a comparison with \teff that is independent of our own in terms of both data and models. 

\begin{figure*}
    \centering
    \includegraphics[width=\textwidth]{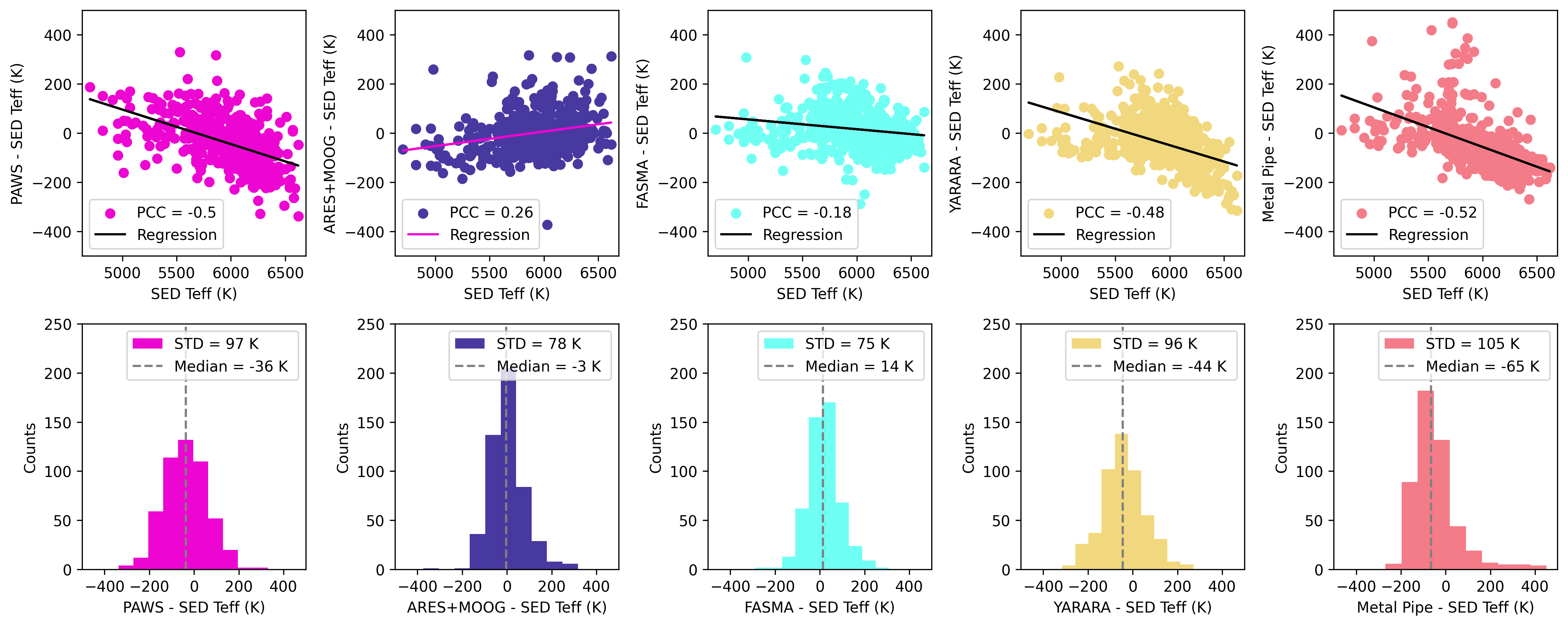}
    \caption{First row : difference between the \teff derived by each method and the \teff from SED fitting, plotted against SED \teff. The Pearson Correlation Coefficient (PCC) and regression lines are additionally included. Second row : Histograms of the differences between spectroscopic and SED \teff, with standard deviation (STD) and median values displayed.}
    \label{fig:sed_teff}
\end{figure*}

A comparison of the \teff from each method in this work to the SED results is shown in Figure \ref{fig:sed_teff}. The top row of Figure \ref{fig:sed_teff} shows the difference between the \teff from each spectroscopic method and that from SED fitting, plotted against the SED \teff, and the bottom row shows the histograms of these differences. Observing the top row, we see that for \textsc{PAWS}, \textsc{YARARA}, and \textsc{Metal Pipe} there is a negative trend between the \teff difference and the SED \teff, with PCCs of -0.49, -0.48, and -0.52, respectively. Restricting to the cooler-than-solar regime, as in Section \ref{pastel}, transforms these correlation coefficients to -0.03, 0.08, and -0.06, respectively. This replicates the observation that the negative correlation drastically reduces, or disappears entirely, when comparing to the PASTEL catalogue in the solar-and-cooler regime in Section \ref{pastel}, suggesting that further calibration for these methods is required for hotter stars.  We again see a positive correlation in difference between \teff results when comparing \textsc{ARES+MOOG} in Figure \ref{fig:sed_teff}, as was observed when comparing to PASTEL, however to a less significant effect with a PCC of 0.26. When we compare the difference between \textsc{FASMA} and SED \teff to the  SED \teff we see a negative correlation, with a PCC of -0.18.  Given the strong agreement between the \textsc{FASMA} \teff, \feh, and \logg and those from PASTEL shown in Section \ref{pastel}, this may reflect an inherent systematic in the comparison of spectroscopic to SED fitting results. We additionally see more negative PCCs when comparing \textsc{PAWS}, \textsc{YARARA}, and \textsc{Metal Pipe} \teff to SED \teff than when comparing to PASTEL. Furthermore, although we again see a positive correlation when comparing the \textsc{ARES+MOOG} results in Figure \ref{fig:sed_teff}, this is to a smaller magnitude at 0.26, rather than 0.39 when comparing to the PASTEL catalogue.

As expected from observing the top row of Figure \ref{fig:sed_teff}, the histograms in the bottom row reveal the difference between spectroscopic and SED \teff have a negative peak for \textsc{PAWS}, \textsc{YARARA}, and \textsc{Metal Pipe}, indicating overall underestimation in \teff compared to SED results from these methods. Despite the trend observed in the top row, the histogram for \textsc{ARES+MOOG} shows a peak close to 0~K, indicating the majority of targets have an \textsc{ARES+MOOG} \teff that agrees well with the SED fitting result. The same is observed for the \textsc{FASMA} histogram, indicating that \textsc{ARES+MOOG} and \textsc{FASMA} are the methods that produce a \teff in the best agreement to SED fitting.

\subsection{Surface gravity}
\label{speclogg}
It is stressed in the literature that spectroscopic \logg often shows biases and offsets \citep[e.g.][]{sozzetti2007, torres2012, mortier2014}. This is, however, dependent on the spectroscopic method, not a broad limitation of spectroscopy itself. With specialised analysis using extensive wavelength coverage and spectral line constraints, spectroscopic methods can recover \logg to within $\approx$ 0.05~dex of asteroseismic values \citep{brewer2015}. The disagreements in \logg determined by the methods in this work reflect the different choices in line lists, model atmospheres, and spectral content employed. Due to this, we selected the 30 targets for which spectroscopic \logg shows the largest disagreement between spectroscopic method results to perform comparisons within this section.

\begin{figure*}
    \centering
    \includegraphics[width=\textwidth]{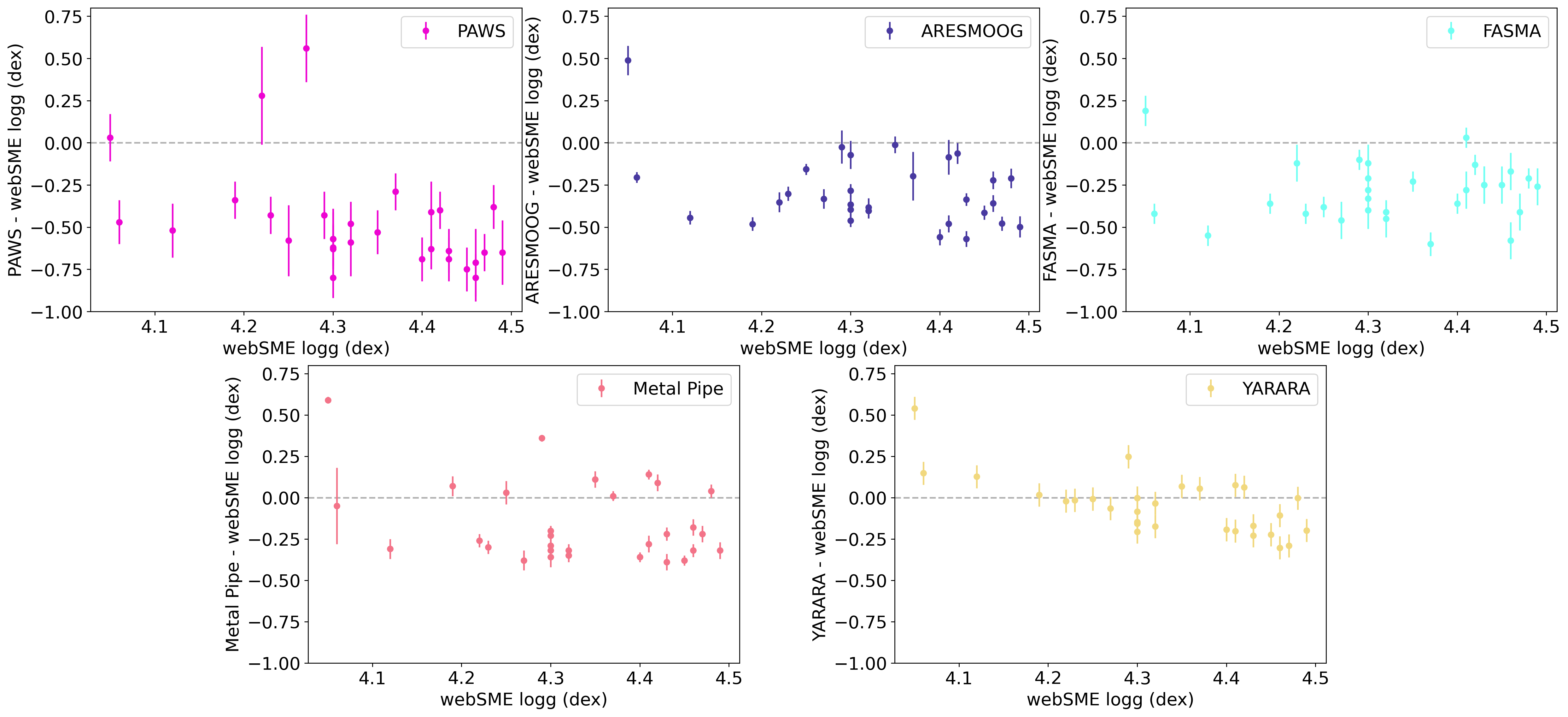}
    \caption{Difference between the \logg derived by each spectroscopic method and that from \textsc{webSME}, plotted against the \textsc{webSME} \logg. This covers the 30 stars with the most discrepant \logg from the different spectroscopic methods.}
    \label{fig:logg_v_sme}
\end{figure*}

\subsubsection{Direct Comparisons of \logg}
In addition to \textsc{PAWS}, \textsc{ARES+MOOG}, \textsc{FASMA}, \textsc{YARARA}, and \textsc{Metal Pipe}, we applied \textsc{webSME} to the 30 \logg discrepant stars. To save computation time we ran only the region 5000 - 6000~{\AA} within \textsc{webSME}, with all targets beginning with solar input parameters. This spectral region includes the Mg I b triplet, known to have high sensitivity in the wings to \logg via pressure broadening \citep{valenti_fischer2005, torres2012}. Given that \textsc{webSME} directly computes \logg from this region, we ran it on all 30 stars in this section to provide an additional, homogeneously derived spectroscopic comparison.

In Figure \ref{fig:logg_v_sme}, for each method we have plotted the difference in \logg when compared to the \textsc{webSME} value against the \textsc{webSME} \logg for each of the 30 stars.  It is clear from this that \textsc{PAWS} underestimates \logg significantly for these targets when compared to the \textsc{webSME} results, with the median difference from \textsc{webSME} \logg being -0.57~dex, far exceeding the median uncertainty of 0.13~dex. To a lesser degree, however still with significance, \textsc{ARES+MOOG}, \textsc{FASMA}, and \textsc{Metal Pipe} also underestimate \logg compared to \textsc{webSME}, with median differences of -0.34, -0.28, and -0.24~dex, respectively. The \logg values from \textsc{YARARA} differ from \textsc{webSME} results by a median value of -0.05~dex, indicating better agreement on average, however we observe a systematic trend in the relationship between the difference in \logg and the \textsc{webSME} value itself. Splitting this sample into the lower half of \textsc{webSME} \logg ($\leq$ 4.32), we see a median difference between \textsc{YARARA} and \textsc{webSME} \logg of -0.02 dex. For the higher \logg half, this median difference increases to -0.18 dex.

\subsubsection{Forward Modelling}
To further assess the accuracy of the spectroscopic \logg values returned by each method, we used webSME's forward modelling capabilities to compare synthesised spectra based on the results to the observed spectra for all 30 of the discrepant \logg stars. The forward modelling takes \teff, \logg, \feh, \vmic, \vmac, and \vsini\,as inputs for synthesising the spectra. Multiple options are available for the line list and solar composition employed in the synthesis -- we opted for the \textit{Gaia-ESO (Y,Y|U) atomic} linelist \citep{heiter2021} and solar reference composition from \citet{asplund2021}. We note that this line list was optimised using the solar composition from \citet{grevesse2007}. The small scale differences introduced by using a different solar composition here, however, do not impact the relative comparisons between methods that we perform. All forward modelling was performed within the region 5000 - 6000\,{\AA}, containing the Mg I b triplet, rather than the whole spectrum, to save computation time.  Due to the sensitivity of these lines to \logg, inaccurate \logg values would be strongly visible in the synthesised lines not representing the observed ones.

For results from methods in which not all broadening parameters (\vmic, \vmac, and \vsini) are determined, we used the inverse variance weighted mean from \textsc{PAWS}, \textsc{FASMA}, and \textsc{webSME} results. \textsc{ARES+MOOG} provides \vmic for all targets, hence the mean was only used for \vmac and \vsini, whereas all three mean broadening parameters were required for both \textsc{Metal Pipe} and \textsc{YARARA}.

\begin{figure*}
    \centering
    \includegraphics[width=\textwidth]{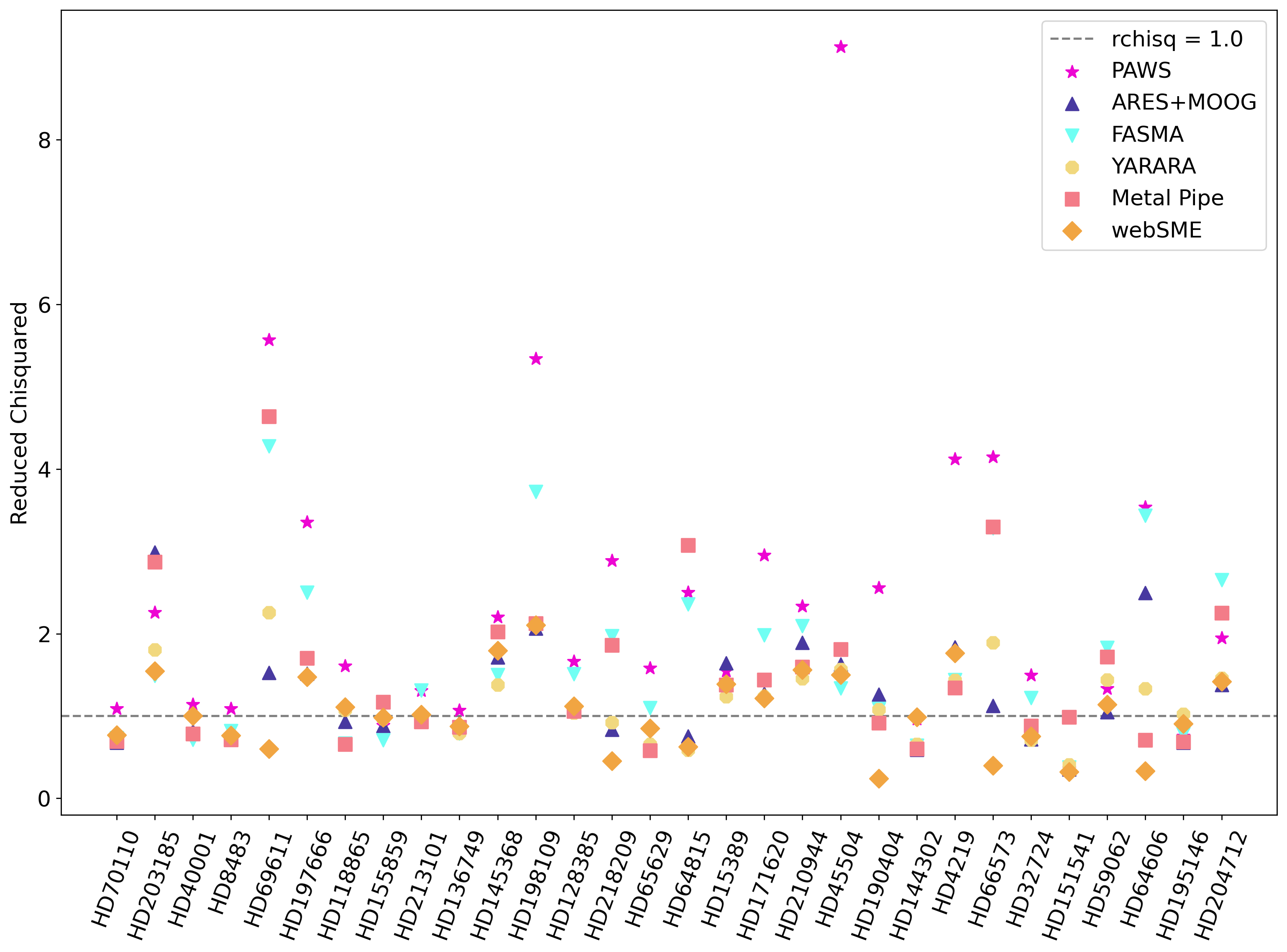}
    \caption{Reduced chisquared of each forward-modelled spectrum when compared to the observed one, using the results from each method. This covers all 30 discrepant \logg targets.}
    \label{fig:forward_model}
\end{figure*}

We used a reduced chisquared metric to directly compare how well the synthesised spectra match the observed one. The result of this is shown in Figure \ref{fig:forward_model}, with the median reduced chisquared from each code shown in Table \ref{tab:median_nrc}. 
The nature of the reduced chisquared is that a value of 1.0 fits the data in accordance with its uncertainties, a value less than 1.0 is overfitting, and a value larger than one indicates a poorer fit. As seen in Table \ref{tab:median_nrc}, \textsc{webSME} produces synthetic spectra fitting best to the observed ones around the Mg I b triplet, whereas \textsc{PAWS} performs poorly in this region, confirming the likely largest inaccuracy in the \logg determined by \textsc{PAWS}.

\begin{figure}
    \centering
    \includegraphics[width=\linewidth]{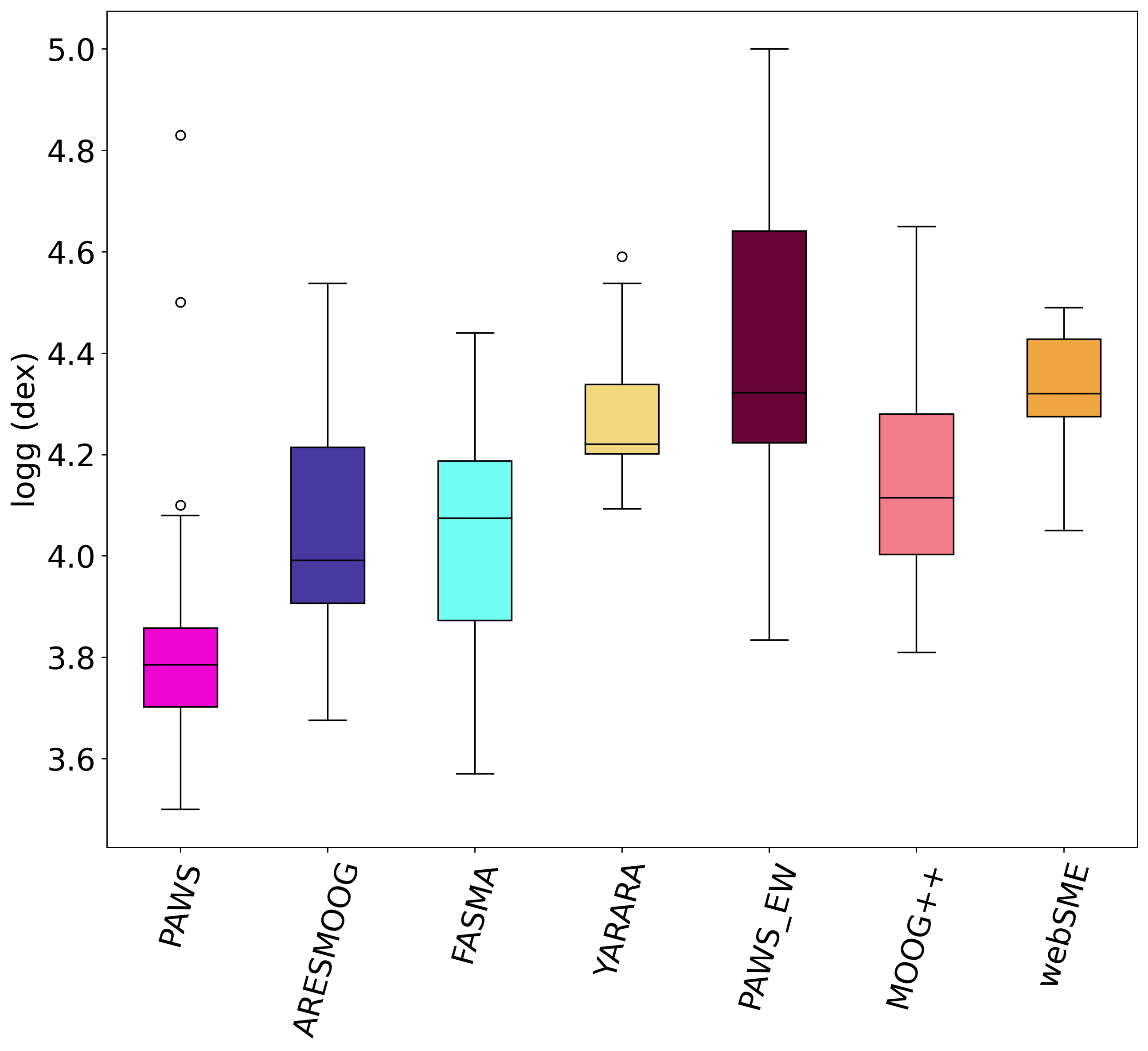}
    \caption{Box-and-whisker plot of the \logg values determined by each code for the sample of 30 discrepant \logg targets, in addition to the \logg for these produced by only the equivalent widths section of \textsc{PAWS}. Open circles represent \logg values that extend beyond the minimum or maximum extremes of the data.}
    \label{fig:logg_hist}
\end{figure}

Given that we see median reduced chisquared values close to 1.0 in Table \ref{tab:median_nrc} for the EW based methods (\textsc{ARES+MOOG} and \textsc{YARARA}) we would also expect the EW section of \textsc{PAWS} to produce a well-fitting \logg value. Figure \ref{fig:logg_hist} displays a box-and-whisker plot of the \logg values from each code in this sample of 30 stars, along with the \logg produced by the EW section of \textsc{PAWS}. Although displaying a much larger range of \logg values, the results from \textsc{PAWS} EW align far more with those from \textsc{webSME} than the full \textsc{PAWS} results, indicating these may be a more reliable measurement. 

\begin{table}
    \centering
    \caption{Median chisquared from comparing synthesised spectra generated with the output of a specific code to the observed spectrum.}
    \label{tab:median_nrc}
    \begin{tabular}{|c|c|}
    \hline
        Code & Median reduced chisquared \\
        \hline
        \textsc{PAWS} & 1.805 \\
        \textsc{ARES+MOOG} & 1.057 \\
        \textsc{FASMA} & 1.408 \\
        \textsc{YARARA} & 1.076 \\
        \textsc{Metal Pipe} & 1.255 \\
        \textsc{webSME} & 0.995 \\
        \hline
    \end{tabular}
\end{table}

\subsubsection{Isochrone \logg}

\begin{figure*}
    \centering
    \includegraphics[width=\textwidth]{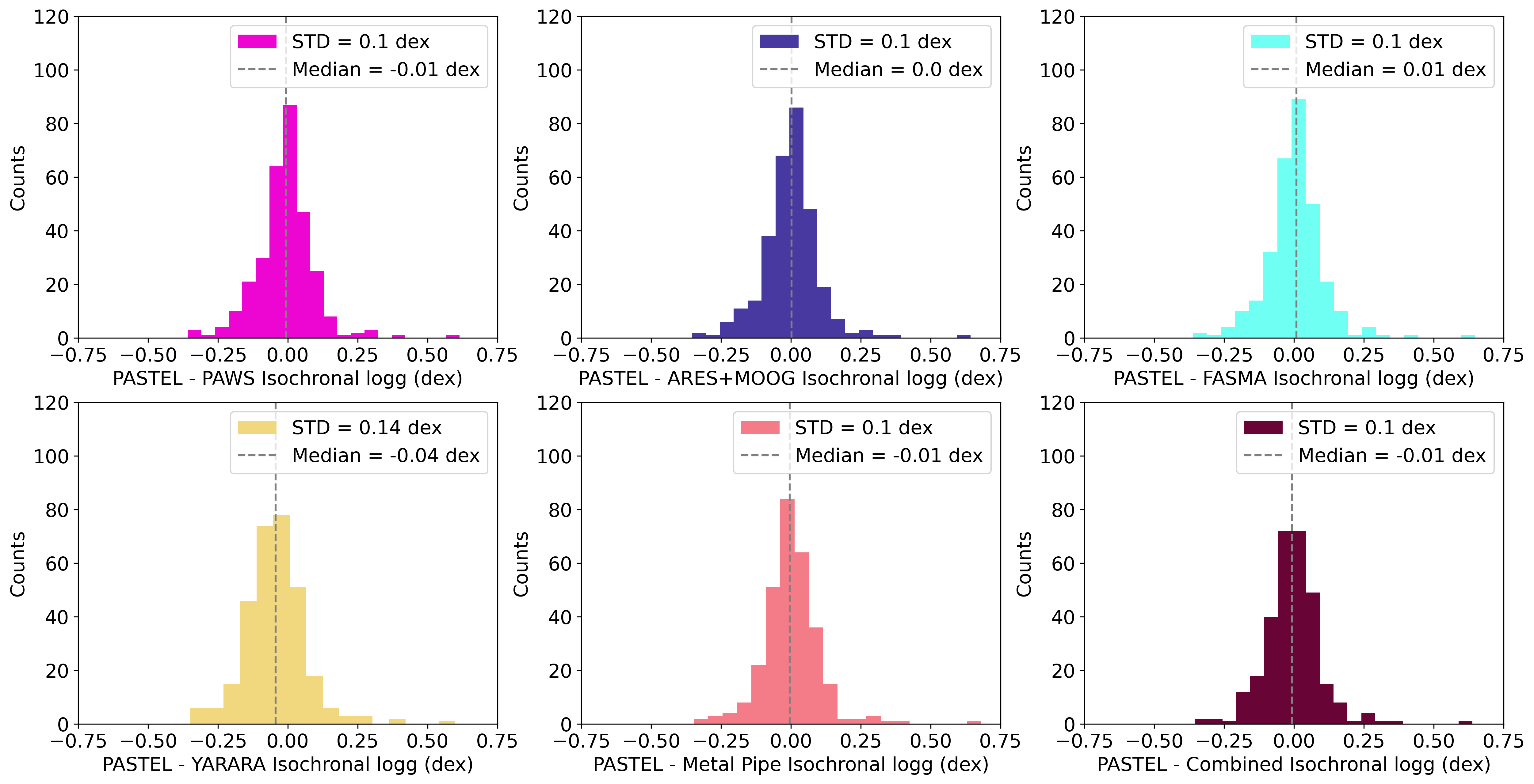}
    \caption{Histograms of the differences between the isochronal \logg for each method and the PASTEL \logg, with standard deviation (STD) and median shown for each distribution.}
    \label{fig:iso_logg}
\end{figure*}

As \textsc{isochrones} determines mass and radius, we can also obtain \logg values from this method. As with Figure \ref{fig:big_pastel_comp}, we can again compare to the PASTEL catalogue, allowing systematic trends to be made apparent. This comparison is shown in Figure \ref{fig:iso_logg}, with the results of inputting the results from each code to \textsc{isochrones} and the combined isochrone posteriors shown through histograms of their difference from the PASTEL \logg. Using these isochrone \logg values, the trends of Figure \ref{fig:big_pastel_comp} have been largely mitigated. Within the context of this sample and the spectroscopic methods tested, this indicates that the combined isochrone posteriors provide the most consistent set of \logg values. We do, however, note that radii determined through isochrones, and hence \logg, are significantly model dependent and may not agree with empirical measurements \citep[e.g.][]{boyajian2015}.

\begin{figure}
    \centering
    \includegraphics[width=0.5\textwidth]{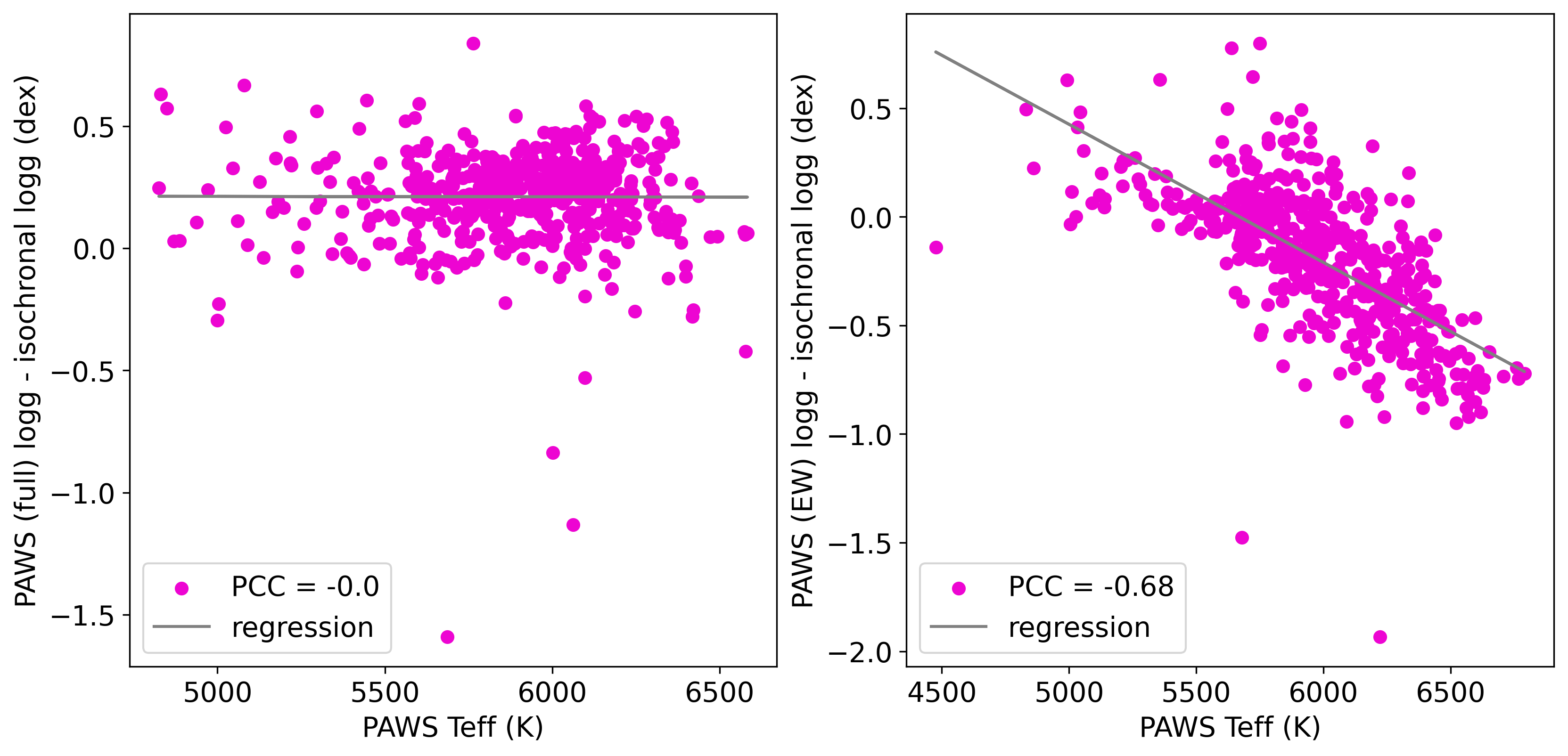}
    \caption{Left panel : Difference between the \logg from the full \textsc{PAWS} pipeline and from the combined isochronal posteriors, plotted against \textsc{PAWS} \teff. Right panel : Difference between the \logg from the EW section of \textsc{PAWS} and the combined isochronal posteriors, plotted against the \textsc{PAWS} \teff.  Both panels additionally show the Pearson Correlation Coefficient (PCC) of the data, in addition to the fitted linear regression lines.}
    \label{fig:ew_synth_logg}
\end{figure}

We additionally investigate whether the use of the equivalent width or synthesis methods influences the observed underestimation of \logg. We show results from this in Figure \ref{fig:ew_synth_logg}. The left hand panel shows the difference between the full \textsc{PAWS} spectroscopic \logg and the combined isochronal \logg, plotted against the \textsc{PAWS} \teff. The right hand panel repeats this, albeit comparing the \teff from only the EW of \textsc{PAWS}, rather than the full pipeline. Although large scatter is seen in both plots, there is a striking negative correlation between the difference in \logg from EW and the stellar effective temperature, replicating the results shown by \citet{mortier2014}. Given this correlation, and the large scatter we see in spectroscopic \logg values throughout Section \ref{speclogg}, we conclude that the combined isochrone \logg values provide the most self-consistent surface gravities from this sample. It is important, however, to remain aware that \logg values derived from isochrones are model dependent, and retain the influence of other stellar parameters \citep{bonaca2012, boyajian2015}. 

\subsection{Microturbulent Velocity \vmic}
An important consideration in our analysis is the influence of spectroscopically determined \vmic on other derived atmospheric parameters. For this investigation, comparison is limited to the codes that provide \vmic, being \textsc{ARES+MOOG}, \textsc{FASMA}, and \textsc{PAWS}.

\begin{figure*}
    \centering
    \includegraphics[width=\textwidth]{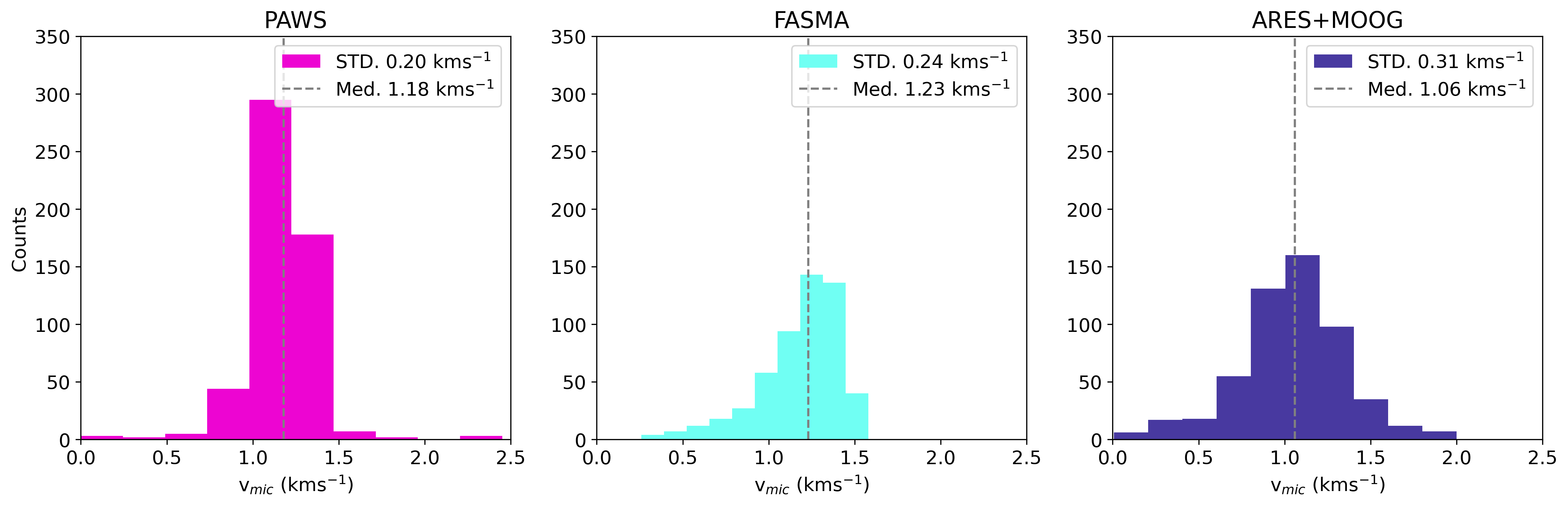}
    \caption{ Histograms of the determined microturblent velocity (\vmic) from \textsc{PAWS}, \textsc{FASMA}, and \textsc{ARES+MOOG}. Each histogram also notes the standard deviation (STD) and median (Med.) values from each code.}
    \label{fig:vmic_hist}
\end{figure*}

Figure \ref{fig:vmic_hist} displays the histograms of the resulting \vmic determined by each respective code, in addition to the median and standard deviation of each sample. From this, we see all three methods clustering in a region of \vmic $\sim$ 1.0 - 1.2 km s$^{-1}$. This is consistent with expectations for a sample of FGK dwarfs \citep[e.g.][]{Tsantaki2013, mortier2014, magrini2022}, indicating a general physical agreement between the methods.

There are, however, evident systematic differences between the methods. The median \vmic values across this common stellar sample vary by up to $\sim$ 0.17 km s$^{-1}$, with \textsc{FASMA} producing the highest median \vmic of 1.23 km s$^{-1}$ and \textsc{ARES+MOOG} producing the lowest of 1.06 km s$^{-1}$. These offsets are on a comparable scale to the internal scatter in the individual methods, highlighting that \vmic is indeed sensitive to the choice of methods and models.

A difference in the dispersions of the resulting \vmic distributions is also apparent, with \textsc{PAWS} showing the lowest standard deviation ($\sigma$ = 0.20 km s$^{-1}$), and \textsc{ARES+MOOG} showing the largest ($\sigma$ = 0.31 km s$^{-1}$). These differences in how tightly \vmic is constrained could be attributed to a variety of reasons, potentially reflecting choices in line lists, models, methodology, or parameter degeneracies. 

Overall, the differences in \vmic distributions from each method reflect method-dependent systematics of $\sim$ 0.1 - 0.3 km s$^{-1}$, exceeding the median uncertainties of 0.06 km s$^{-1}$, 0.03 km s$^{-1}$, and 0.06 km s$^{-1}$ for \textsc{PAWS}, \textsc{FASMA}, and \textsc{ARES+MOOG} respectively. This is further evidenced by the median RMSD in \vmic across the three methods being 0.10 km s$^{-1}$. Given the degeneracies between \vmic and other spectroscopic parameters, these systematics may contribute to the differences observed between \teff, \logg, and \feh in the methods, however are unlikely to be the sole cause.

\begin{figure*}
    \centering
    \includegraphics[width=\linewidth]{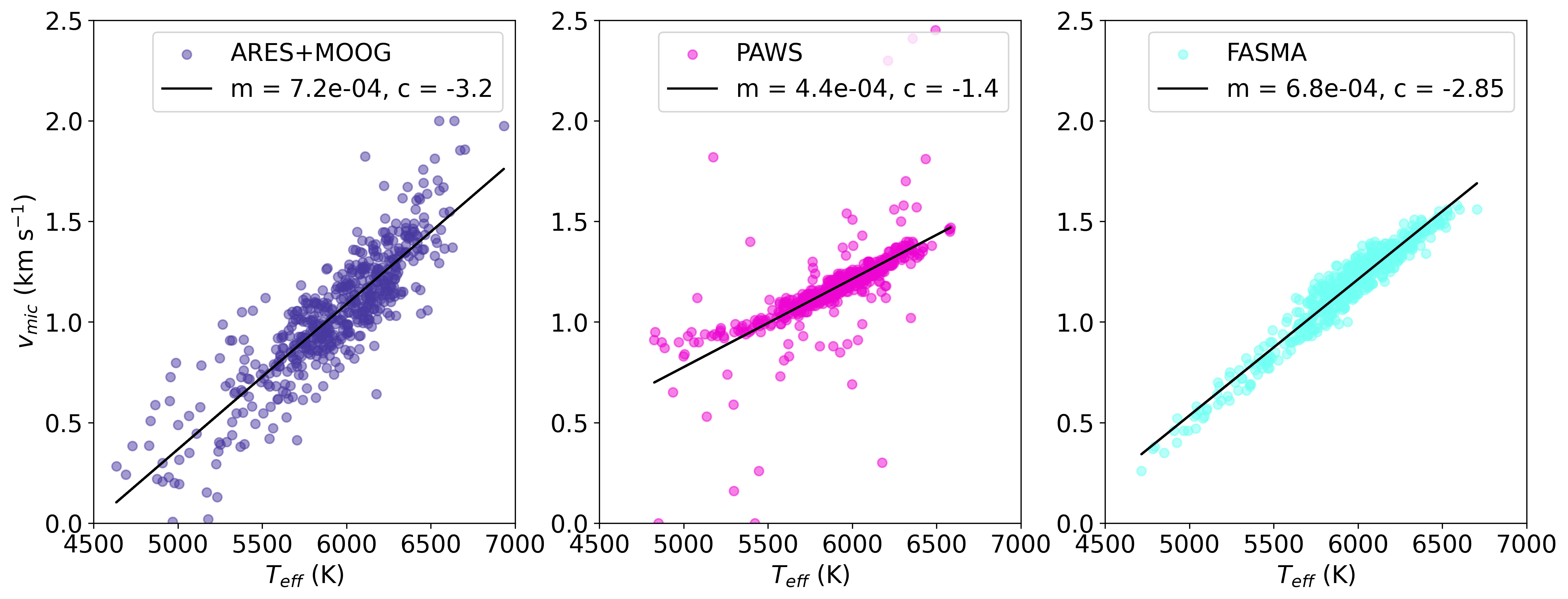}
    \caption{The relationship between \vmic and \teff for \textsc{ARES+MOOG} (left), \textsc{PAWS} (centre), and \textsc{FASMA} (right). The slope (m) and intercept (c) of the fitted linear regression lines are shown in the legend of each subplot.}
    \label{fig:vmic_teff}
\end{figure*}

In addition to the overall \vmic distributions, we also observe a clear positive trend between \vmic and \teff across all three methods, shown in Figure \ref{fig:vmic_teff}. This is consistent with expectations for FGK dwarfs, for which higher \teff is associated with higher velocity convective motions. 

Although this positive trend is broadly followed for all three methods, Figure \ref{fig:vmic_teff} shows that the exact relationship differs. This indicates that while the understood physical relationship between \vmic and \teff is concretely recovered, its absolute calibration remains sensitive to the choice of method.

\subsection{Stellar Radius}
\begin{figure*}
    \centering
    \includegraphics[width=\textwidth]{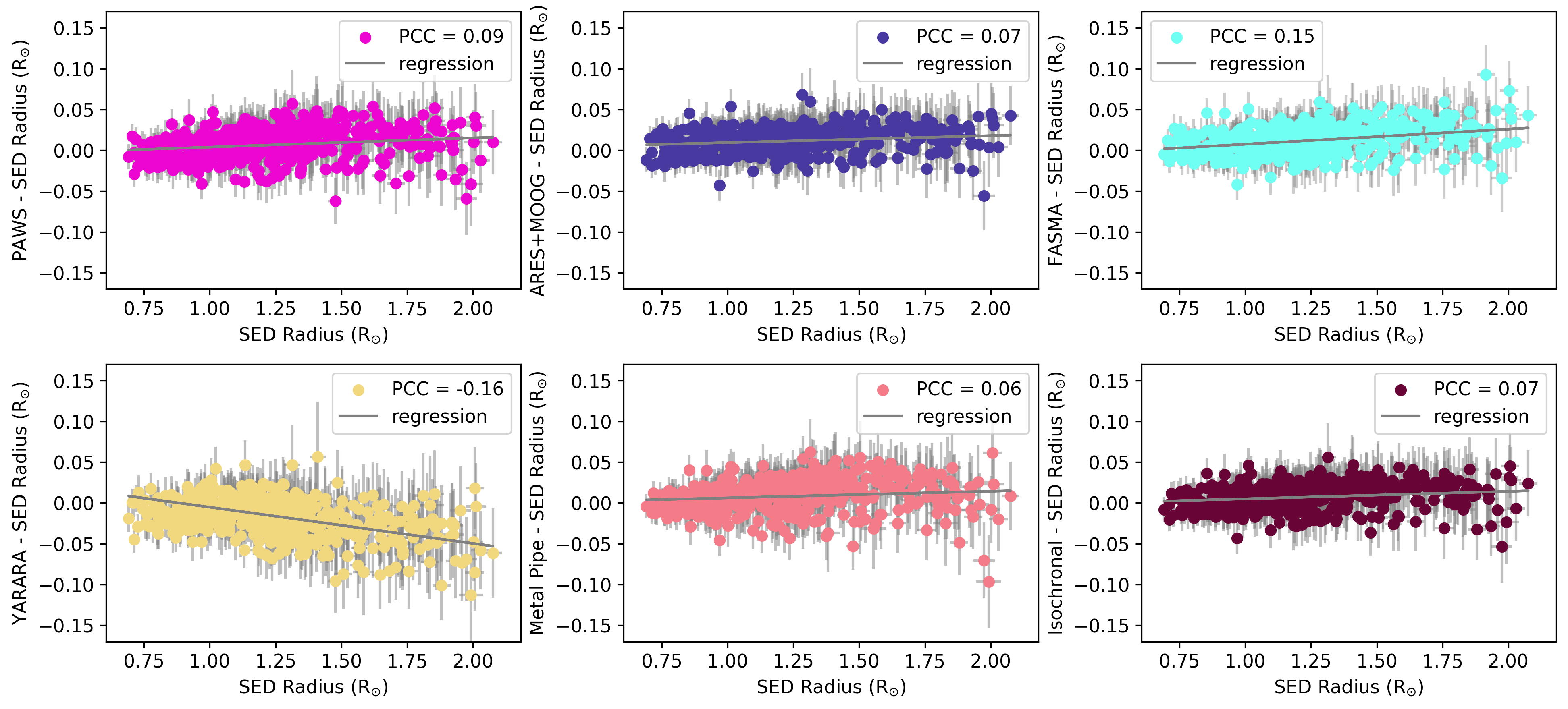}
    \caption{Differences in radii from each spectroscopic method + \textsc{isochrones}, and the radii from SED fitting, plotted against SED radius for each method. The lower right hand panel shows the results from combining all \textsc{isochrones} results. Four SED outliers have been removed from all of these plots. The Pearson Correlation Coefficient (PCC) and linear regression lines fitted to the data are shown in each panel.}
    \label{fig:radius_sed}
\end{figure*}

As previously described, we obtained stellar radii from SED fitting as part of GR8-1. In Figure \ref{fig:radius_sed}, we show the difference between the radii from using each spectroscopic method plus \textsc{isochrones}, as described in Section \ref{isochrone_method}, and the radii from SED fitting, plotted against the SED fitting radii. In all subplots, we see a slight upwards trend, with the difference between spectroscopic+\textsc{isochrones} and SED radii increasing with the SED radius.

An excellent agreement between the values is seen, with the RMSD between spectroscopy+\textsc{isochrones} and SED fitting being under 0.10~R$_{\odot}$ for all sets of results. Figure \ref{fig:radius_sed} also shows the results from the combined posterior, shown as the `Isochronal' results in the bottom right hand panel. From combining all spectroscopic+\textsc{isochrones} results, we observe an RMSD of 0.08~R$_{\odot}$ from the SED radii. 
 
\section{Impact on Exoplanetary Parameters}
\label{exoplanets}
In the context of exoplanet characterisation or demographic studies, it is important to understand the effect that the scatter in stellar parameters has on exoplanetary parameters. Here, we consider the planetary radius, mass, and equilibrium temperature.

In terms of the effect on exoplanetary radii, R$_{p}$, we can use the transit depth
\begin{equation}
    \mathrm{Depth} = \left(\frac{R_{p}}{R_{\star}}\right)^{2}
\end{equation}
to investigate the effect of scatter in stellar radius. Considering only the uncertainty induced by the scatter in stellar radius, we see an uncertainty in planetary radius of
\begin{equation}
    \label{radius_unc}
    \sigma_{R_{p}} =  \sigma_{R_{\star}} \sqrt{D}
\end{equation}
where $\sigma_{R_{\star}}$ is uncertainty in stellar radius, and $\sigma_{R_{p}}$ is the resultant uncertainty in planetary radius. We calculate the scatter in radius as the RMSD of the individual results from the median value. Across all targets, the median scatter in radius is 0.02~R$_{\odot}$. Investigating the Earth-Sun system, we see an uncertainty in planetary radius of 0.02 R$_{\oplus}$ (2.00 \%) induced purely by the median scatter in stellar radius. As we extend to the cooler regime of K dwarfs, we use the case study of HD219134 \citep{motalebi2015} -- a K dwarf with 4 orbiting planets. Looking at HD219134 b, we see an induced uncertainty in planetary radius of R$_{\oplus}$ (2.57 \%). We additionally test on CoRoT-11b \citep{gandolfi2010}, a planet orbiting at 0.0436~AU around an F dwarf with $R$ = 1.37~$R_{\odot}$ and \teff = 6440~K. In this system, we see an uncertainty in planetary radius purely from scatter in stellar radius of 0.23~R$_{\oplus}$ (1.46 \%). For HD219134 b and CoRoT-11 b, the quoted uncertainties in planetary radii are 0.09~R$_{\oplus}$ and 0.34~R$_{\oplus}$, respectively.  From the NASA Exoplanet Archive \citep{christiansen2025}, the median fractional uncertainty on exoplanetary radius in the literature is 7.36\%, higher than uncertainty predicted to arise from scatter in stellar radius alone. It is important to note, however, that this prediction comes from scatter due to differing spectroscopic models only -- the effects of varying evolutionary models using \textsc{isochrones} is not explored here.

To determine the significance of the effect of stellar mass scatter on the exoplanetary mass, we use the equation
\begin{equation}
    M_{p} \sin i \approx K\left(\frac{P}{2\pi G}\right)^{\frac{1}{3}}M_{\star}^{\frac{2}{3}}
\end{equation}
where M$_{p}$ is planetary mass, P is planetary orbital period, G is the gravitational constant, and M$_{\star}$ is the stellar mass. We assumed a circular orbit and a planetary mass that is  negligible compared to the stellar mass. Isolating the contribution to uncertainty of scatter in stellar mass, we obtain the following equation to determine uncertainty in planetary mass
\begin{equation}
    \sigma_{Mp} \approx \frac{2}{3} \frac{\sigma_{M\star}}{M_{\star}}M_{p}
\end{equation}
where $\sigma_{Mp}$ and $\sigma_{M\star}$ are the uncertainties in planetary and stellar mass, respectively. Again taking the median scatter as RMSD from the median of the combined isochronal posteriors, we find this to be 0.05~M$_{\odot}$ in stellar mass. For the Earth-Sun system, this scatter induces an uncertainty of 0.03~M$_{\oplus}$ (3.00 \%) in planetary mass. For HD219134 b, with a mass of 4.36 $\pm$ 0.44~M$_{\oplus}$, a scatter of 0.05~M$_{\odot}$ would contribute an uncertainty of 0.19~M$_{\oplus}$ (4.36 \%). CoRoT-11 b, with a mass of 740.47 $\pm$ 108.05~M$_{\oplus}$, would have an uncertainty contributed from stellar mass scatter of 18.01~M$_{\oplus}$ (2.43 \%).  Typical uncertainties in planetary mass appear to be significantly higher than these fractional uncertainties due to stellar mass scatter -- see \citet{mortier2018} (20.95 \%), \citet{hoyer2021} (7.25 \%), \citet{mantovan2024} (9.71 \% \& 40.68 \%), \citet{schmerling2025} (8.79 \%).

Taking all results from this work for the targets analysed, we calculate the scatter on each target as the RMSD from the median value. In \teff, the median scatter across all targets is 76~K. Taking the situation of an Earth-Sun twin, with all orbital and physical parameters equal to that of the Earth and Sun, we used the following equation to calculate planetary equilibrium temperature
\begin{equation}
    T_{\textrm{eq}} = T_{\textrm{eff}} \sqrt{\frac{R}{2a}}(1-A_{B})^{\frac{1}{4}}
    \label{eqt}
\end{equation}
where  $R$ is stellar radius, $a$ is orbital semi-major axis, and $A_{B}$ is the Bond albedo of the planet. When propagating the uncertainty due to scatter in stellar \teff from using different methods, we consider that stellar radius is often not directly inputted into equation \ref{eqt}, rather $\frac{R_{\star}}{a}$ is generally found from transit information, and assume this to be the case for this purpose.  Using the NASA Exoplanet Archive \citep{christiansen2025}, we find the median fraction uncertainty in $\frac{R_{\star}}{a}$ from transit measurements to be 0.16. We can use the following equation to translate the scatter in stellar \teff from the use of different methods to a fractional uncertainty in the planetary $T_{\textrm{eq}}$;
\begin{equation}
    \frac{\sigma_{T_{\textrm{eq}}}}{T_{\textrm{eq}}} = \sqrt{(\frac{\sigma_{T_{\rm eff}}}{T_{\rm eff}}) + \frac{1}{4}R_{\rm frac}^2  }
    \label{teq_uncert}
\end{equation}
where $R_{\rm frac}$ is the fractional uncertainty in $\frac{R_{\star}}{a}$, which we set to 0.16 as per the median value in the literature. From equation \ref{teq_uncert}, all three case studies as previously investigated return a fractional uncertainty of 0.04 in the planetary equilibrium temperature. This suggests a noise floor of at least 4\% fractional uncertainty should be present in all T$_{\textrm{eq}}$ measurements. In recent literature, planetary equilibrium temperatures appear to often have uncertainties in the range 2.0\% - 3.5\% (e.g. \citet{mortier2018} -- 2.13\%, \citet{hoyer2021} -- 3.10\%, \citet{mantovan2024} -- 2.96\% \& 2.98\%, \citet{scott2025} -- 2.40\%, \citet{schmerling2025} -- 3.27\%), suggesting that the noise floor introduced by scatter in \teff is not fully represented in the current state of the field.

From these investigations, it is strongly suggested that the scatter in spectroscopic stellar parameters from differing model choices is not a dominating factor in exoplanetary mass and radius uncertainties. However, as previously stated, this comparison does not reflect the scatter that can also be introduced by the use of differing evolutionary models in the determination of mass and radius. For planetary equilibrium temperature, the noise floor from the combination of scatter in \teff and median uncertainty in $\frac{R_{\star}}{a}$ is higher than typical uncertainties in the literature, suggesting uncertainties in T$_{\textrm{eq}}$ are generally underestimated. Additionally, as the precision and accuracy of exoplanet instrumentation and methodology increases, the noise floors investigated in this work will become increasingly prevalent and require careful attention. 

\section{Conclusions}
\label{conc}
By providing uniformly formatted, high resolution, high SNR spectra, the \texttt{gr8stars} catalogue forms a uniquely useful aid in studying the systematic effects of different spectroscopic methods. This work shows the results of applying five spectroscopic methods, differing in methodology, models, and assumptions, to the same data for a subset of 585 \textsc{gr8stars} targets.  

For the stellar effective temperature, surface gravity, and metallicity, we show that the precision errors often quoted for spectroscopic parameters are far outweighed by the scatter that comes from the use of different methods. Given that we consistently see disagreements between results at a significance larger than the precision uncertainties, we can see that spectroscopic atmospheric parameters are still influenced heavily by the disagreement between methodologies and models, thus require uncertainties reflective of this. From this work, suitable uncertainties for spectroscopically derived \teff, \logg, and \feh require inflations of at least 76 K, 0.14 dex, and 0.07 dex respectively.

As part of this work, we also study the spectroscopic \logg using \textsc{webSME} for a direct comparison of 30 stars. As expected, we see large scatter in the spectroscopic \logg returned from each method. We demonstrate that the isochronal \logg, which incorporates information from photometry and parallax, has improved agreement with literature values from the PASTEL catalogue. Due to this, we conclude that the isochronal \logg should be relied upon, rather than the spectroscopic one. 

We also investigate the masses, radii, and ages of the stars, using isochronal fitting with the spectroscopic effective temperature and metallicity as part of the input. We propagate the scatter induced in these parameters from using \teff and \feh from different spectroscopic methods to exoplanetary equilibrium temperature, radius, and mass. We find that the induced scatter does not outweigh the typical fractional uncertainties on exoplanetary mass and radius in the literature, whereas fractional uncertainties in planetary equilibrium temperature present a noise floor above those typically found in the literature.

While we found that the scatter in planetary radii and masses induced by different spectroscopic inputs is typically lower than uncertainties reported in the literature, it is important to recognise that this represents only part of the total uncertainty sources. In this work, stellar mass, radius, and age were determined using a single set of stellar evolution models. As a result, the noise floors that we estimate due to scatter do not account for systematic uncertainties resulting from model choice. 

In addition, the demands of next-generation exoplanet surveys and studies require increasingly precise stellar properties. When we look particularly at the detection and characterisation of Earth-sized planets, stellar properties require uncertainties at a level for which the scatter explored in this work becomes significant.

While we focus on FGK dwarfs within this work, the expansion of this to cooler dwarfs would be particularly valuable. Spectroscopic modelling becomes increasingly difficult for stars with lower effective temperatures, and these are important targets for upcoming exoplanet surveys, including \textit{PLATO}.

Future work should aim to extend this analysis with the incorporation of multiple stellar evolution models and grids. This will allow the full quantification of systematic uncertainties in derived stellar and planetary properties. As observational precision continues to improve, a thorough understanding of these systematics will be essential in reaching the full potential of next-generation exoplanet science.

\section*{Acknowledgements}
We are grateful to the anonymous referee for their constructive feedback that improved the quality of this work.

AVF acknowledges the support of the IOP through the Bell Burnell Graduate Scholarship Fund.

A.M. acknowledges funding from a UKRI Future Leader Fellowship, grant number MR/X033244/1 and a UK Science and Technology Facilities Council (STFC) small grant ST/Y002334/1.

The Flatiron Institute is a division of the Simons Foundation.

This work benefited from support from the European Research Council (ERC) under the European Union’s Horizon 2020 Framework Programme (grant agreement no. 865624).

This project has received funding from the European Union’s Horizon 2020 research and innovation programme under grant agreement No 101008324 (ChETEC-INFRA).

This project has received funding from the European Research Council (ERC) under the European Union’s Horizon 2020 research and innovation programme (CartographY GA. 804752)

JIGH and ASM acknowledge financial support from the Spanish Ministry of Science, Innovation and Universities (MICIU) project PID2023-149982NB-I00.

S.G.S. acknowledge support from FCT through FCT contract nr. CEECIND/00826/2018 and POPH/FSE (EC)

NCS is co-funded by the European Union (ERC, FIERCE, 101052347). Views and opinions expressed are however those of the author(s) only and do not necessarily reflect those of the European Union or the European Research Council. Neither the European Union nor the granting authority can be held responsible for them. This work was supported by FCT - Fundação para a Ciência e a Tecnologia through national funds by these grants: UIDB/04434/2020 DOI: 10.54499/UIDB/04434/2020, UIDP/04434/2020 DOI: 10.54499/UIDP/04434/2020.

This research has made use of the NASA Exoplanet Archive, which is operated by the California Institute of Technology, under contract with the National Aeronautics and Space Administration under the Exoplanet Exploration Program.

C.A.W. would like to acknowledge support from the UK Science and Technology Facilities Council (STFC, grant number ST/X00094X/1)
\section*{Data Availability}
All stellar parameters derived through this work will be available through Vizier CDS. Spectra are available upon request to the authors.



\bibliographystyle{mnras}
\bibliography{gr8stars2} 




\appendix
\section{PASTEL \teff, \feh, and \logg trends}
\label{srsly_name_this_properly}
\begin{figure*}
    \centering
    \includegraphics[width=\textwidth]{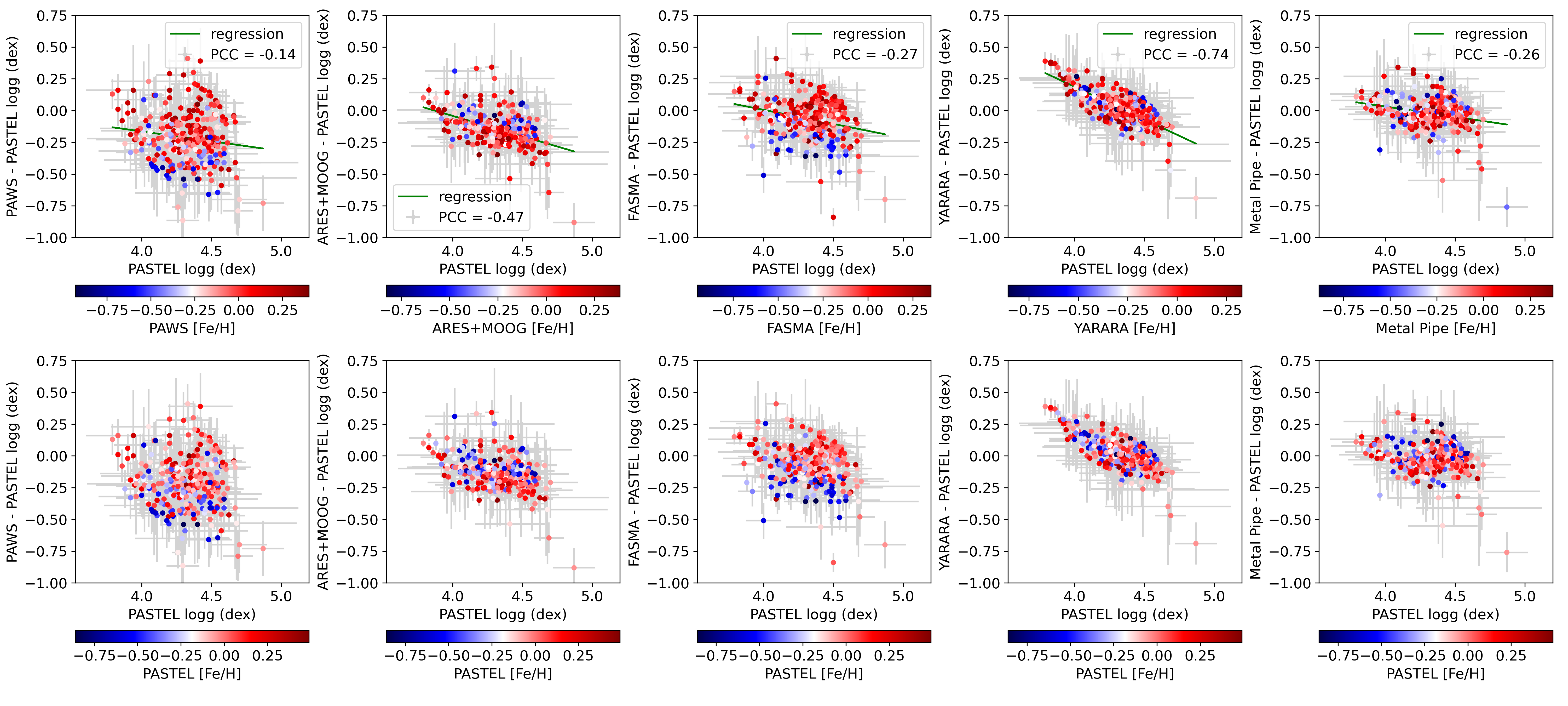}
    \caption{Difference between \logg from this work and PASTEL, coloured by the \feh from each method on the top row, and the PASTEL \feh on the second row. The top row shows the Pearson Correlation Coefficient (PCC) and linear regression lines fitted to the data.}
    \label{fig:logg_feh_col}
\end{figure*}

\begin{figure*}
    \centering
    \includegraphics[width=\textwidth]{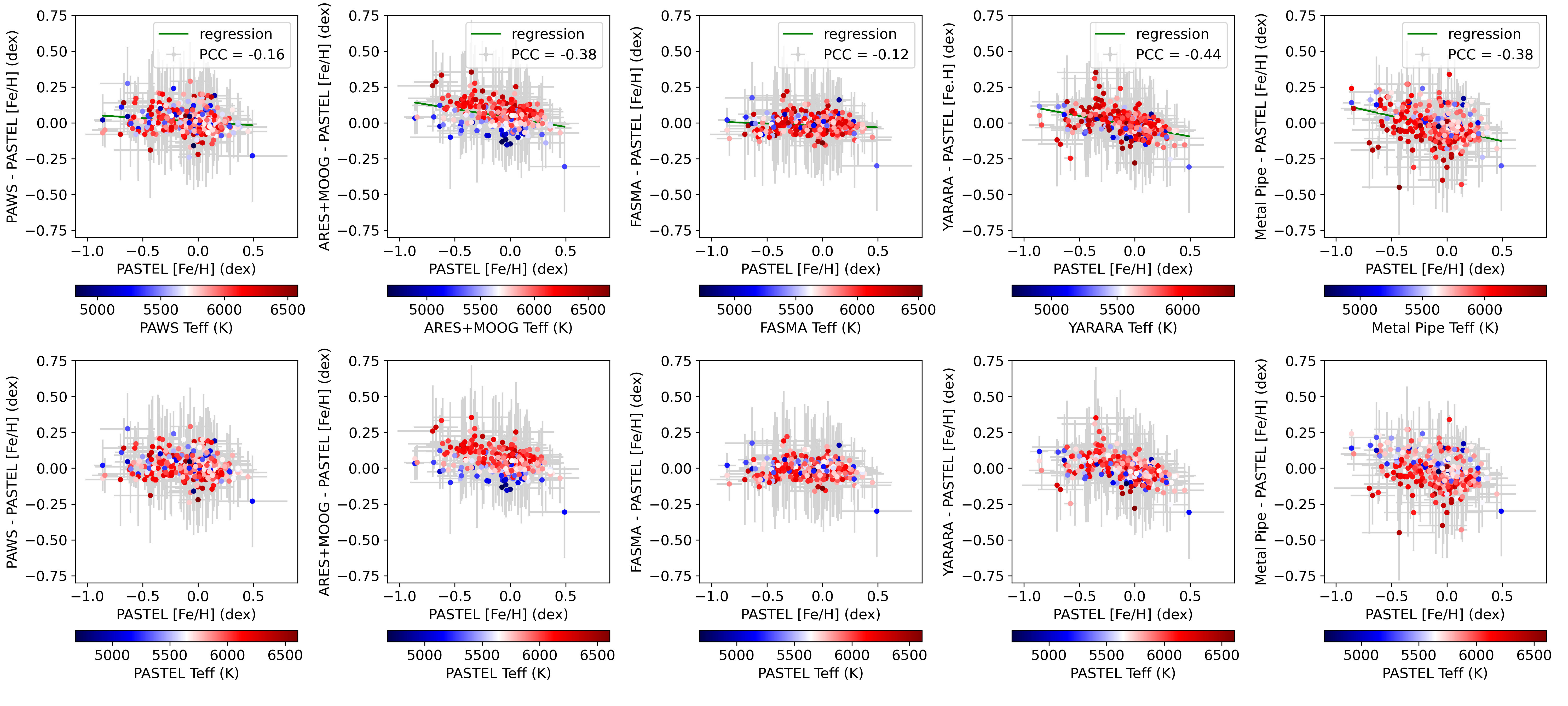}
    \caption{Difference between \feh from this work and from PASTEL, coloured by the \teff from each method on the first row, and the PASTEL \teff on the second row.  The first row shows the Pearson Correlation Coefficient (PCC) and linear regression lines fitted to the data.}
    \label{fig:feh_teff_col}
\end{figure*}

Figures \ref{fig:logg_feh_col} and \ref{fig:feh_teff_col} display the differences between values determined spectroscopically in this work, and those from the PASTEL catalogue. Figure \ref{fig:logg_feh_col} compares \logg from this work to PASTEL, coloured by the \feh, whereas Figure \ref{fig:feh_teff_col} compares \feh to PASTEL, coloured by \teff.

\bsp	
\label{lastpage}
\end{document}